\documentclass[journal]{IEEEtran}
\usepackage{ifpdf}
\usepackage{cite}

\ifCLASSINFOpdf
\usepackage[pdftex]{graphicx}
\graphicspath{{../pdf/}{../jpeg/}{pic/}}
\DeclareGraphicsExtensions{.pdf,.jpeg,.png}
\else
\usepackage[dvips]{graphicx}
\graphicspath{{../eps/}}
\DeclareGraphicsExtensions{.eps}
\fi

\usepackage{amsmath}
\usepackage{algorithmic}
\usepackage{array}
\usepackage{pgfplots}

\usepackage[font=normalsize,labelfont=sf,textfont=sf]{caption}
\usepackage[font=normalsize,labelfont=sf,textfont=sf]{subcaption}

\usepackage{lmodern}  
\usepackage{mathrsfs} 

\usepackage{url}
\usepackage{amssymb}  
\usepackage{csquotes} 

\usepackage[ruled,linesnumbered]{algorithm2e}
\usepackage{mathtools} 
\usepackage{cases} 
\usepackage{empheq} 
\usepackage[capitalize]{cleveref} 
\usepackage{cuted} 
\usepackage{bm} 
\usepackage{amsthm} 
\usepackage{tabulary,booktabs} 
\usepackage{tensor}
\usepackage{multirow}

\newtheorem{theorem}{Theorem}

\newtheorem{lemma}{Lemma}

\theoremstyle{definition}
\newtheorem{definition}{Definition}

\newcommand{\bb}[1]{\mathbb{#1}}
\newcommand{\mb}[1]{\mathbf{#1}}
\newcommand{\mcal}[1]{\mathcal{#1}}

\newcommand{\cond}{i\in\mcal{I},j\in\mcal{L}_i}
\newcommand{\C}{C(J,\bb{R}^L)}
\newcommand{\bx}{\bm{x}}
\newcommand{\eq}[1]{Eq.~(\ref{#1})}
\newcommand{\fig}[1]{Fig.~\ref{#1}}

\newcommand{\sect}[1]{Section~\ref{#1}}

\newcommand{\tn}{\tilde{n}}
\newcommand{\tk}{\tilde{k}}

\newcommand{\norm}[1]{\left\lVert#1\right\rVert}

\usepackage[]{xcolor} 
\definecolor{text}{HTML}{EBDCB2}
\definecolor{bg}{HTML}{282828}
\colorlet{paper}{red!33!green!33!blue!33}
\colorlet{grey}{white!10!black}
\usepackage{listings}

\begin{document}

\title{Dynamics in Coded Edge Computing for IoT: A Fractional Evolutionary Game Approach}

%

\author{Yue~Han,
Dusit~Niyato, \IEEEmembership{Fellow, IEEE},
Cyril~Leung, 
Chunyan~Miao, 
Dong~In~Kim, \IEEEmembership{Fellow, IEEE}
\thanks{
Y. Han is with Alibaba Group and the Alibaba-NTU Joint Research Institute (JRI), Nanyang Technological University (NTU), Singapore. E-mail: hany0028@e.ntu.edu.sg.}
\thanks{D. Niyato is with the School of Computer Science and Engineering (SCSE), NTU, Singapore. E-mail: dniyato@ntu.edg.sg.}
\thanks{C. Leung is with The University of British Columbia (UBC) and the Joint NTU-UBC
Research Centre of Excellence in Active Living for the Elderly (LILY). E-mail: cleung@ntu.edu.sg.}
\thanks{C. Miao is with SCSE, NTU, Singapore, the Alibaba-NTU JRI, and LILY, Singapore. E-mail: ascymiao@ntu.edu.sg.}
\thanks{Dong In Kim is with the Department of Electrical and Computer Engineering, Sungkyunkwan University, Suwon, South
Korea. E-mail: dikim@skku.ac.kr.}
}

%
%

\markboth{Journal of \LaTeX\ Class Files,~Vol.~14, No.~8, August~2015}%
{Shell \MakeLowercase{\textit{et al.}}: Bare Demo of IEEEtran.cls for IEEE Journals}
%



\maketitle

\begin{abstract} 

%
%
%

%
%
%
%
%
%
%
Recently, coded distributed computing (CDC), with advantages in intensive computation and reduced latency, has attracted a lot of research interest for edge computing, in particular, IoT applications, including IoT data pre-processing and data analytics. Nevertheless, it can be challenging for edge infrastructure providers (EIPs) with limited edge resources to support IoT applications performed in a CDC approach in edge networks, given the additional computational resources required by CDC. 
In this paper, we propose \enquote{coded edge federation}, in which different EIPs collaboratively provide edge resources for CDC tasks. To study the Nash equilibrium, when no EIP has an incentive to  unilaterally alter its decision on edge resource allocation, we model the coded edge federation based on evolutionary game theory. Since the replicator dynamics of the classical evolutionary game are unable to model economic-aware EIPs which memorize past decisions and utilities, we propose \enquote{fractional replicator dynamics} with a power-law fading memory via Caputo fractional derivatives. The proposed dynamics allow us to study a broad spectrum of EIP dynamic behaviors, such as EIP sensitivity and aggressiveness in strategy adaptation, which classical replicator dynamics cannot capture. 
Theoretical analysis and extensive numerical results justify the existence, uniqueness, and stability of the equilibrium in the fractional evolutionary game. The influence of the content and the length of the memory on the rate of convergence is also investigated. 

%

\end{abstract}

%
\IEEEpeerreviewmaketitle

\section{Introduction}

{\IEEEPARstart{C}{oded} distributed computing (CDC) \label{sec:intro} \cite{leeSpeedingDistributedMachine2018a,reisizadehCodedComputationHeterogeneous2017,yuPolynomialCodesOptimal2017,tandonGradientCodingAvoiding2017,ravivGradientCodingCyclic2019,yuLagrangeCodedComputing2019,liCodedTeraSort2017,liCodedComputingMitigating2020,ngComprehensiveSurveyCoded2021} has recently received much attention in both machine learning and information theory domains for its advantages in speeding up computations for large parallelable tasks. 
In parallel computing, a task is divided into several sub-tasks, each of which is assigned to a worker. Thus, the overall computation time is determined by the slowest worker. If some workers take significantly longer to complete their sub-tasks, the overall performance of parallel computing is degraded. 
To combat such a straggler effect, CDC treats delayed workers as erasures and leverages coding techniques to effectively create redundant computational tasks. 
In this way, the results of a subset of workers whose completion times are relatively shorter can be used to recover the original results. 
The idea of CDC has been extensively studied when resources are plentiful, e.g., in centralized data centers. Examples of CDC used to mitigate straggler effects include gradient descent computation in large-scale machine learning platforms \cite{tandonGradientCodingAvoiding2017,ravivGradientCodingCyclic2019}, 
matrix multiplication \cite{leeSpeedingDistributedMachine2018a,reisizadehCodedComputationHeterogeneous2017,yuPolynomialCodesOptimal2017}, and data sorting \cite{liCodedTeraSort2017}. It is also applied to address security and privacy issues \cite{yuLagrangeCodedComputing2019} and to improve communication bottlenecks \cite{liCodedComputingMitigating2020}. More details are provided in \cite{ngComprehensiveSurveyCoded2021}. 

Recently, there has been a growing interest in applying CDC to edge computing scenarios, such as federated learning \cite{prakashCodedComputingLowLatency2021,haCodedFederatedComputing2019} and task offloading for latency-sensitive applications\cite{kimCodedEdgeComputing2020,hanOpportunisticCodedDistributed2021}. 
This is due to the fact that prevalent edge computing tasks such as data pre-processing, data analytics, object detection, and $3$D rendering are parallelable, computation-intensive, and latency-sensitive \cite{kimCodedEdgeComputing2020}. In addition, computation results transmitted over noisy wireless networks are error-prone. 

However, a direct implementation of CDC at the edge is challenging. First, the objective of using CDC is to improve the Quality-of-Service (QoS) of latency-critical applications, e.g., IoT applications, in which data are collected and processed at the edge. Therefore, the placement of the computing resources (e.g., CPU, memory, and storage disk) is critical, and a shorter distance to the cloud user with lower transmission delays and bandwidth costs is preferable.A cloud computing approach \cite{wangCloudComputingPerspective2010}, in which computation is transferred to centralized data centers, incurs high transmission delays. Alternatively, geographically diversified edge clouds, or sometimes called distributed clouds \cite{rashidDistributedCloudComputing2018,agarwalVolleyAutomatedData2010,alicherryNetworkAwareResource2012,endoResourceAllocationDistributed2011}, consisting of a large number of small data centers, geographically spread at the edge and interconnected by medium to high speed links \cite{rashidDistributedCloudComputing2018}, are closer to the cloud users and provide more location diversity. The close proximity benefits real-time applications in which latency is important (e.g., virtual reality, gaming, machine-to-machine communication, industrial applications, interactive collaboration, smart grid control) and laws require user data to be stored in a specific locality, such as the same county, city, or office \cite{coadyDistributedCloudComputing2015}. A challenge with this approach is that the computation capacity of a single edge cloud is considerably smaller compared to a centralized cloud. This will degrade the performance of CDC, as it trades additional computational resources for a shorter completion time. 
Meanwhile, multiple edge infrastructures providers (EIPs) may place edge resources at the same location (\cref{systemfigure}), or relatively close to each other in some regions \cite{caoEdgeFederationIntegrated2020}. Thus, an opportunity for federation among edge clouds controlled by different EIPs is promising. 

\begin{figure}
\centering
\centering
\includegraphics[width=0.98\linewidth]{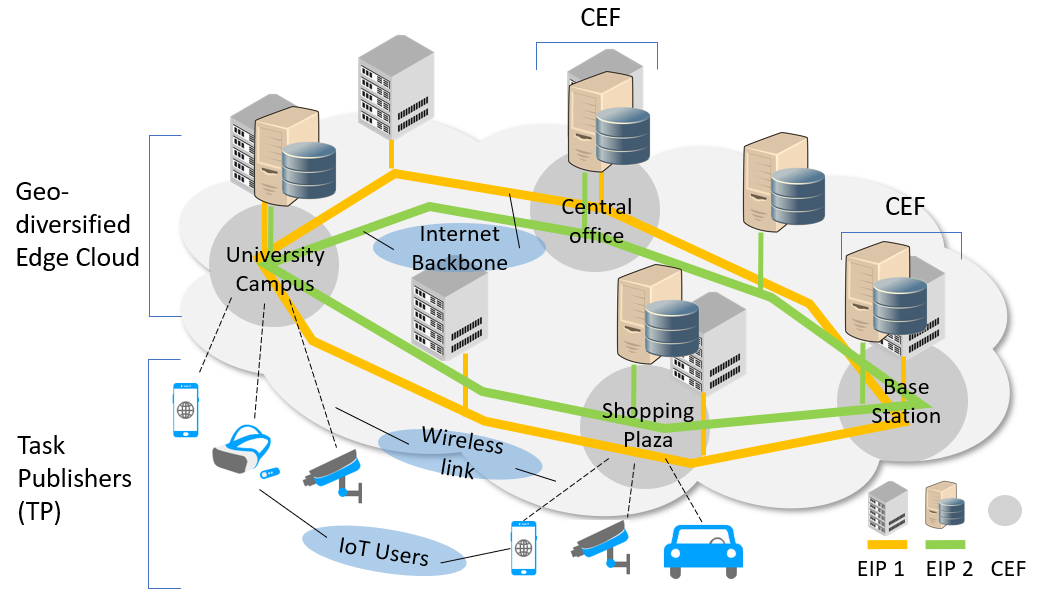}
\caption{System model of the coded edge federation (CEF)}
\label{systemfigure}
\label{fig:system}
\end{figure}

To address the aforementioned challenge, 
%
%
we propose {a framework called} \enquote{coded edge federation} (CEF), in which multiple EIPs cooperate and share their edge resources. 
When edge resources from one EIP are insufficient, nearby edge clouds belonging to different EIPs can collaboratively deliver the required CDC services. The benefits of the CEF are as follows. First, CEF's close proximity feature helps to avoid the long transmission delays caused by offloading to far-away centralized clouds or distant clusters located in other regions. Second, CEF can unify the heterogeneous edge resources from various EIPs by enforcing EIPs to contribute standardized resources measured by a set of workers, each corresponding to a standard set of resources, e.g., CPU, memory, and storage disk. Third, to ensure the performance of CDC, particularly when normal service requests are heavy at the edge, a CDC task is given a higher priority and served ahead of normal services. Fourth, with more artificial intelligence (AI) and IoT services deployed at the edge, CEF allows an EIP to rapidly respond to the increasing QoS requirements of demanding applications by enabling an EIP to supplement its limited edge capacity from its federation partner EIPs, before its new edge clouds are in place.

Existing studies of resource management for CDC \cite{liFundamentalTradeoffComputation2018,ngGametheoreticApproachCollaborative2021,kimIncentiveBasedCodedDistributed2020} focus on either the resource trade-off between computation and communication or the incentive mechanism for edge devices (e.g., smart phones and drones) to cooperate as a coalition and deliver latency-sensitive IoT services. There have been few studies on a federation among EIPs to provide CDC services in edge networks. We aim to address this problem using the framework of a non-cooperative game in which EIPs, each with a large number of geo-diversified edge clouds, behave as selfish players. 
EIPs need to decide on the amount of resources (number of workers) to contribute from their edge clouds to the CEF. 
Because priority is given to CDC tasks by the CEF, the service quality for other regular edge services offered by the same workers will be affected. 
Thus, the EIP needs to balance the resources allocated by it and other players to the federation. 
After many rounds of strategic interactions, a balanced and stable strategy profile is achieved whereby no EIP has an incentive to unilaterally change its decision. 
This equilibrium may be a desirable solution for the EIPs that participate in the CEF.

We adopt an evolutionary game approach, a particular type of non-cooperative game characterized by bounded rationality and a revision protocol \cite{weibullEvolutionaryGameTheory1997}, to identify such an equilibrium point. In addition, we consider incorporating a power-law fading memory \cite{tarasovaConceptDynamicMemory2018} of past decisions into a well-known type of revision protocol, namely replicator dynamics, via the left-sided Caputo fractional derivatives \cite{kilbasTheoryApplicationsFractional2006}. The proposed new dynamics are thus referred to as \textit{fractional replicator dynamics}, and the associated evolutionary game is called \textit{fractional evolutionary game}. Fractional replicator dynamics can overcome the myopic nature of classical replicator dynamics \cite{tarasovaConceptDynamicMemory2018} and capture the economically rational nature of EIPs who act as business owners and use past memory to improve their decision-making in the market. The memory parameter $\alpha$ in Caputo fractional derivatives can capture additional and important aspects of EIP decision-making process, e.g., the content and the length of the memory, which reflect an EIP's sensitivity and aggressiveness in the strategy adaptation.

We summarize the major contributions of the paper as follows:
\begin{itemize}
	\item We highlight the edge resource insufficiency problem in implementing CDC at the edge and propose the \textit{coded edge federation} (CEF) framework based on evolutionary game to address the challenge.

\item We propose \textit{fractional replicator dynamics} with a power-law fading memory via fractional calculus to model strategy adaptation of economic-aware EIPs, which overcome the myopic limitation of classical replicator dynamics used in an evolutionary game approach. 

\item 
We theoretically show and experimentally verify the validity of the proposed dynamics, by establishing the existence, uniqueness, and stability of the solution. In addition, the proposed dynamics demonstrate faster convergence rates compared to the classical ones.

\end{itemize}

{The rest of this paper is organized as follows. \cref{sec related work} reviews related works. \cref{sec:system-model} presents the system model and the problem formulation as a classical evolutionary game. \cref{sec:fracgame} proposes the fractional replicator dynamics and formulates a fractional evolutionary game. \cref{section: equiblirum anlaysis} provides a theoretical study of the game equilibrium in terms of its existence, uniqueness, and stability. \cref{sec: performance evalutation} investigates the dynamic behavior of EIPs experimentally. Finally, \cref{sec conclusion} summarizes the main findings and lists potential future work.

\section{Related Work} 
\label{sec related work}
\subsection{Coded Distributed Computing (CDC)} 
CDC is a promising technique for addressing the straggler problem in distributed computing \cite{reisizadehCodedComputationHeterogeneous2017,leeSpeedingDistributedMachine2018a,yuPolynomialCodesOptimal2017}. The study in \cite{leeSpeedingDistributedMachine2018a} proposed using erasure codes (e.g., repetition and MDS code) to speed up distributed computing in homogeneous settings. 
In \cite{reisizadehCodedComputationHeterogeneous2017}, the study was extended to heterogeneous settings where different workers receive different amounts of local data. A hierarchical coded computation scheme is proposed in \cite{parkHierarchicalCodingDistributed2018} to utilize the partial results of stragglers. Privacy aspects of coded computing are investigated in \cite{bitarMinimizingLatencySecure2020}. Other studies \cite{tandonGradientCodingAvoiding2017,ravivGradientCodingCyclic2019} propose methods to improve gradient computing for a large-scale machine learning platform. All these studies assume ample computing resources are available, e.g., in a large data center. They do not consider the limited computational resources typically available in edge computing scenarios. 
Recent studies  \cite{ngGametheoreticApproachCollaborative2021,kimIncentiveBasedCodedDistributed2020} focus on applying CDC in edge computing. However, they mainly focus on incentive mechanisms for heterogeneous edge devices. 
In contrast, we study the dynamics of edge resource federation among EIPs which provide coded edge computing for IoT services. In particular, we adopt a fractional evolutionary game approach to model the dynamic behavior of an EIP with memory.

\subsection{Edge Federation}

For edge federation, studies on cross-cloud cooperation architectures, such as Joint Cloud \cite{wangJointcloudCrosscloudCooperation2017} and Hybrid Cloud \cite{meylerMicrosoftHybridCloud2017}, propose to horizontally integrate public and private cloud resources allowing EIPs to elastically handle short-term spikes, e.g., Black Friday for Amazon and Double $11$ for TaoBao. They provide insight into the horizontal edge federation for coded distributed computing. On the other hand, vertical integration in terms of content caching and computation offloading are proposed in  \cite{dziyauddinComputationOffloadingContent2019,shanMultilevelOptimizationFramework2020}, by which frequent content duplication can be reduced \cite{danDynamicContentAllocation2014} and the power and computation limitation of an edge device can be mitigated. In \cite{caoEdgeFederationIntegrated2020}, the integration of edge cloud from both horizontal and vertical perspectives is studied. 
The key differences between our work and existing studies on the edge (or cloud) federation are as follows. First, we mainly consider a horizontal integration among independent EIPs, as the vertical offloading to the cloud may incur potential transmission delay in the back-haul network, which negates the benefits of using CDC to reduce latency for IoT services. 
Second, instead of using a centralized method, which is static and has a high communication overhead, we adopt a dynamic and decentralized resource allocation strategy based on evolutionary game theory. In particular, the rewards competition among multiple EIPs is modeled by fractional replicator dynamics with awareness of past decisions.

\subsection{Power-law Fading Memory and Fractional Calculus}
The concept of memory is widely used in mathematical modeling in various fields,  
e.g., physical sciences \cite{koellerApplicationsFractionalCalculus1984,colemanGeneralTheoryFading1968}, economics  \cite{teyssiereLongMemoryEconomics2006,tarasovaElasticityEconomicProcesses2016,tarasovaConceptDynamicMemory2018}, and control \cite{podlubnyFractionalDifferentialEquations1998}.
In many systems, the dynamics depend not only on the information at the present time but also the past. Fractional calculus has been widely used to represent past memory  \cite{kilbasTheoryApplicationsFractional2006,teyssiereLongMemoryEconomics2006}. 
In \cite{tarasovaConceptDynamicMemory2018}, a Volterra operator and power-law fading memory are used to aggregate past information via the left-sided Caputo derivatives. The extension to memory-aware elasticity via fractional calculus is investigated in \cite{tarasovaElasticityEconomicProcesses2016}. The authors in \cite{tarasovaEconomicGrowthModel2017}  extends the economic growth model under the effect of dynamic memory with a constant pace. In coded edge federation, EIPs as market players are economically rational in the sense that they can recall past information and use it in future decision making. It is therefore necessary to model EIP memory effects in the strategy adjustment. 
However, classical replicator dynamics are known to be myopic and memoryless \cite{tarasovaConceptDynamicMemory2018}. One of our main contributions is to study the strategies adaptation of memory-aware EIPs in the coded edge federation. 

\label{sec system model}

\section{System Description} \label{sec:system-model}
In this section, we first introduce the system model, notation, and assumptions in \cref{sub:system}. Then, we formulate the evolution of the CEF as an evolutionary game in \cref{sub:game}.

\begin{table*}[]
	\caption{Notation used in the system model}
\begin{tabular}{|p{0.12\linewidth}|p{0.84\linewidth}|}
		\hline
		Notation                                   & Explanation                                                                                                                                                                                                                                                      \\ \hline
		$\mcal{I}, i, I$                           & set   of EIPs, an EIP, and the number of EIPs in the   system model                                                                                                                                                                   \\ \hline
		$\mcal{E}_i, e_i,   E_i$                   & set   of edge clouds, an edge clouds, and the total number of clouds for EIP $i$                                                                                                                                                           \\ \hline
		$\mcal{P},p,P$                             & set   of types of coded tasks, a type of coded task, and the total number of types of tasks                                                                                                         \\ \hline
		$(n_p,k_p,D_p,\lambda_p)$                  & code configuration, task size, and task frequency for a type $p$ task                                                                                                                                                  \\ \hline
		$\mcal{L}_i,l_i,L_i$. &        set of pure strategies (possible contributions) for EIP $i$, a pure strategy, and the maximum contribution                         \\ \hline
		$l,\bm{l},L$                               & total number of contribution to an CEF by all EIPs, a pure-strategy profile, and the dimension of the pure-strategy profile. $l=\sum_{i\in\mcal{I}}l_i$; $\bm{l}=[l_i]_{i\in\mcal{I}}$; $L=\sum_{i\in\mcal{I}}L_i +I$ \\ \hline
		$N_{i,l_i}$                                & estimated number   of edge clouds of EIP $i$ contributing $l_i$ workers to the coded edge   federation                                                                                                                                                                    \\ \hline
		$\bm{x}_i=[x_{i,l_i}]_{l_i\in \mcal{L}_i}$        & mixed strategy for an edge cloud of EIP  $i$                                                                                                                                                                                                                                                                                                                                                                                                                       \\ \hline
$\bm{x}=[\bm{x}_i]_{i\in\mcal{I}}$  &mixed strategy  profile of all EIPs         \\ \hline
		$\Delta_i,\Theta$                          & space   of $\bm{x}_i$ and the space of $\bm{x}$. $\Delta_i\in\bb{R}^{L_i+1},   \Theta=\times_{i\in\mathcal{I}}\Delta_i$                                                                                                                                          \\ \hline
		$\tn_i,\tk_i,i\in\mcal{I}$                            & number of workers of EIP $i$ among the set of $n_p$ workers requested by a type $p$ task and among the set of $k_p$ workers used to recover the result $\sum_{i\in\mcal{I}}\tn_i=n_p$;$\sum_{i\in\mcal{I}}\tk_i=k_p$     \\ \hline
		$p(\tn_i;n_p,\bm{l},\bm{\tn})$                               & probability that $\tn_i$ workers of EIP $i$ are among the set of  $n_p$ workers requested by a type $p$ task                                                                                                                                                   \\ \hline
		$p(\tk_i;k_p,\bm{\tn},\bm{\tk})$                               & probability that $\tk_i$ workers of EIP $i$ are among the set of  $k_p$ workers used to recover the original result                                                                                                                                   \\ \hline
		$R^{(0)}_i,R^{(1)}_i,R^{(2)}_i$            & base   reward, additional reward, and fixed reward for EIP $i$, respectively                                                                                                                                                                                     \\ \hline
		$r_0$, $r_1$, $r_2$                        & unit   reward for  base reward,  additional reward, and fixed reward,   respectively                                                                                                                                                                             \\ \hline
		$C_i^{(1),p}$                              & energy   cost for EIP $i$ for a type $p$ task                                                                                                                                                                                                               \\ \hline
		$C_{i,l_i}^{(2)}$                          & resource   utilization cost for an edge cloud of EIP $i$ contributing $l_i$ workers to an CEF                                                                                                                                              \\ \hline
		$w_i$                                      & resource   utilization, i.e., the fraction of edge resources used by the coded edge   federation for EIP $i$                                                                                                                                                     \\ \hline
		$W_i$, $C_i$, $c_i$                                        & maximum   edge capacity,   fixed cost  to be calibrated for EIP $i$, and cost per CPU cycle for an edge cloud of EIP $i$                                                                                                        \\\hline                                                                                 
		$\pi_i^{(p)}(l_i;\bm{l})$                  & utility that an edge cloud of EIP $i$ contributing $l_i$ workers to an CEF can receive given the pure-strategy profile and a type $p$ task                                                                     \\ \hline
		$\pi_i(l_i;\bm{l})      $                         & utility that an edge cloud of EIP $i$ contributing $l_i$ workers can receive given the pure-strategy profile                                                               \\ \hline
		$u_{l_i,\bm{x}}(l_i;\bm{l})$                          & expected utility that an edge cloud of EIP $i$ contributing $l_i$ workers can receive given the mixed strategy profile                                                                                               \\ \hline
		$u_i(\bm{x})$                              & average   utility that EIP $i$ receives with the mixed strategy profile $\bm{x}$ $\bm{x}$                                                                                                                                                              \\ \hline
	\end{tabular}\label{notation}
\end{table*}

\subsection{System Description, Notation, and Assumptions} \label{sub:system}
As shown in \cref{systemfigure}, the system model includes multiple CDC task publishers (TPs), a set of edge infrastructure providers (EIPs), and the coded edge federation (CEF) consisting of edge clouds controlled by various EIPs. TPs, EIPs, and the CEF communicate with a centralized controller. 

\subsubsection{Task Publishers}\label{para:task publisers}
Let $\mcal{P}=\{1,\ldots,p,\ldots,P\}$ be a set of integers representing $P$ types of tasks requested by the TPs. A type $p$ task is described by a tuple $\{(n_p,k_p,D_p,\lambda_p)\}_{p\in\mcal{P}}$, where the code configuration, task size, and task arrival rate are denoted by $(n_p,k_p)$, $D_p$, and $\lambda_p$, respectively. Various codes, such as repetition code \cite{leeSpeedingDistributedMachine2018a}, MDS code \cite{leeSpeedingDistributedMachine2018a}, Lagrange code \cite{yuLagrangeCodedComputing2019}, and Polynomial code\cite{yuPolynomialCodesOptimal2017}, can be used for different CDC applications. 
With the code configuration $(n_p,k_p)$ in a {homogeneous} setting, a type $p$ task is divided into $k_p$ {equal} sub-tasks and encoded into $n_p$ sub-tasks. Here, we consider a master-worker setting, in which a TP acting as a master sends a CDC request with code configuration $(n_p,k_p)$ to its nearby CEF. Then, a set of $n_p$ workers from the CEF compute $n_p$ sub-results in parallel. After the computation, the sub-results are sent back to the TP. In CDC, the first $k_p$ results received by the TP are sufficient to recover the original result.

\subsubsection{EIPs and Edge Clouds}
Let $\mcal{I}=\{1,\ldots,i,\ldots,I\}$ denote a set of $I$ EIPs, which provide a large amount geo-diversified edge clouds, based on which the CEF is formed. To unify the heterogeneous resource provision from various EIPs for executing CDC tasks, the CEF issues a standardized measure of resource contribution to all EIPs by requesting them to contribute resources in the form of a set of workers, each of which offers a standardized amount of computation and storage capacity. Thus, the resources allocated to the CEF by an EIP can be quantified by the number of workers. 
Let $\mcal{E}_i=\{1,\ldots,e_i,\ldots, E_i\}$ be a set of $E_i$ edge clouds controlled by EIP $i$, geographically distributed at the network edge \cite{tongHierarchicalEdgeCloud2016}, e.g., base stations, access points, central offices, and university campus, which can directly receive CDC task requests from TPs via wireless links. 
Here, we study the region in which different EIPs co-place their edge clouds, or more generally, the region in which the distances among the edge clouds of various EIPs are much shorter than the distances to edge clouds of the same EIP placed in other areas. 
For example, in \cref{systemfigure}, we can assume that the distance between the shopping plaza and the campus is long, and thus EIP $1$ edge clouds placed in the campus and the plaza are far away from each other, whereas edge clouds of EIPs $1$ and $2$ placed in the same area, e.g., campus, are much closer to each other. Therefore, the CEF formed around the campus can benefit the campus-located TPs who have latency-critical tasks. 


\subsubsection{Strategies of EIPs}
As introduced previously, EIPs independently decide how much resources (measured by the number of workers) their edge clouds should contribute when collaborating with neighboring edge clouds of other EIPs. Without loss of generality, we assume that the allocation decision spaces for all edge clouds of an EIP are the same\footnote{The setting can be easily extended if the edge clouds have different decision spaces. For example, we may group the edge clouds with the same decision space and indicate the group index in an additional dimension.} and we denote this common decision space by  $\mcal{L}_i=\{0,1,2,\ldots,l_i,\ldots,L_i\},i\in\mcal{I}$, where $L_i$ is the maximum number of workers that can be allocated to the CEF. Here, $l_i$, i.e., a particular allocation chosen from the set $\mcal{L}_i$, is referred to as a \textit{pure-strategy}, and the set $\mcal{L}_i$ is called the \textit{set of pure-strategies} for EIP $i$. We assume that the set $\mcal{L}_i$ is fixed by the EIP in a given period, e.g., one week, but can be changed for different periods. Moreover, EIPs are allowed to use \textit{probabilistic strategies}, i.e., a probability distribution over $\mcal{L}_i$, with which each pure strategy is to be played with a certain probability. Let $\bm{x}_{i}=[x_{i,0},x_{i,1},\ldots,x_{i,l_i},\ldots,x_{i,L_i}]$ denote the mixed strategy for EIP $i$. Therefore, the number of edge clouds that adopt strategy $l_i$ can be estimated by $N_{i,l_i}=E_ix_{i,l_i}$, if $E_i$ is large, which is common in the case of a large amount of geo-diversified edge clouds \cite{haoOnlineAllocationVirtual2017}. To study the strategy profile, i.e., the combined decisions made by all EIPs, we let $\bm{l}=[l_1,l_2,\ldots,\l_I]$ denote the \textit{pure strategy profile} and $\bm{x}=[\bm{x}_1,\ldots,\bm{x}_I]$ to denote the \textit{mixed strategy profile}. 
Since $\sum_{l_i\in\mcal{L}_i}x_{i,l_i}=1$, we let unit simplex $\Delta_{i}\subset \bb{R}^{L_i+1}$ denote the space of the mixed strategy for EIP $i$ and have $\bm{x}_i\in \Delta_{i}$. 
Let $\Theta=\times_{i\in\mathcal{I}}\Delta_i$ denote the Cartesian product of $\Delta_i$. Then, $\Theta$ is a subspace in $\bb{R}^L$, where $L=I+\sum_{i\in\mcal{I}}L_i$. Therefore, $\bm{x}\in\Theta$. With EIPs independently making their decisions, the probability of having a pure strategy profile $\bm{l}=[l_1,\ldots,l_i,\ldots,l_I]$ is $\prod_{i\in\mcal{I}}x_{i,l_i}$.
\subsubsection{Composition of Workers in the CEF}\label{para:prob}
We can quantify the capacity of an CEF, e.g., the one formulated around the campus in \cref{systemfigure}, as  $l=\sum_{i\in\mcal{I}}l_i$, i.e., the sum of allocation by all EIPs from their edge clouds to this CEF. Thus, with the request of $n_p$-workers from type $p$ tasks sent to this CEF, we consider a random assignment 
from the request to the $l$ homogeneous workers in the CEF for simplicity. 
Let $\tilde{n}_i,i\in\mcal{I}$ denote the number of workers contributed by EIP $i$ among the $n_p$ workers, subject to $0\leq \tilde{n}_i \leq l_i$. 
Let $\bm{\tilde{n}}=[\tilde{n}_1,\tilde{n}_2,\ldots,\tilde{n}_I]$ denote the joint placement by all EIPs, which is subjected to $\sum_{i\in\mcal{I}}\tilde{n}_i=n_p$. Then, the probability that $\tilde{n}_i$ workers of EIP $i$ in performing the type $p$ CDC task is 
%
%

\begin{multline}
p(\tn_i;n_p,\bm{l},\bm{\tn})=
\frac{\binom{l_1}{\tn_1}
\ldots\binom{l_{i-1}}{\tn_{i-1}}
\binom{l_{i+1}}{\tn_{i+1}}
\ldots
\binom{l_{I}}{\tn_{I}}
}{\binom{l}{n_p}}\\
\prod_{i\in\mcal{I}}\bb{I}\{\tn_i\geq 0 \}
\prod_{i\in\mcal{I}}\bb{I}\{l_i\geq \tn_i \}
\bb{I}\{\sum_{i\in\mcal{I}}\tn_i=n_p\},
\label{eq:prob1}
\end{multline}
where $\bb{I}\{A\}$ is the indicator function such that $\bb{I}\{A\}=1$ if the event $A$ is true and $\bb{I}\{A\}=0$ otherwise. Note that \cref{eq:prob1} is based on that the pure-strategy profile $\bm{l}$ and the joint placement $\bm{\tn}$ are given. When $I=2$, \cref{eq:prob1} reduces to a hyper-geometric distribution
\begin{multline}
p(\tn_i;n_p,\bm{l},\bm{\tn})=\frac{\binom{l_1}{\tn_1}\binom{l-l_1}{n_p-\tn_1}}{\binom{l}{n_p}}
\bb{I}\{\tn_1\geq 0, n_p- \tn_1\geq 0\}\\
\bb{I}\{l_1\geq \tn_1, l-l_1\geq n_p-\tn_1\}.
\end{multline}

As soon as TP receives the results from $k_p$ workers, its task is completed. Due to the homogeneous setting, each of the $n_p$ workers is equally likely to be among the first $k_p$ workers to return the results. 
Let $\tk_i,i\in\mcal{I}$ denote the number of workers belonging to EIP $i$ that are in the set of $k_p$ workers, subjected to $0\leq \tk_i \leq \tn_i$, and  $\bm{\tk}=[\tk_1,\ldots,\tk_I]$ denote the joint results of all EIPs. Therefore, the probability that $\tk_i$ workers of EIP $i$ are in the set of first $k_p$ workers is
\begin{multline}
p(\tk_i;k_p,\bm{\tn},\bm{\tk})=
\frac{\binom{\tn_1}{\tk_1}
\ldots\binom{\tn_{i-1}}{\tk_{i-1}}
\binom{\tn_{i+1}}{\tk_{i+1}}
\ldots
\binom{\tn_{I}}{\tk_{I}}
}{\binom{n_p}{k_p}}\\
\prod_{i\in\mcal{I}}\bb{I}\{\tk_i\geq 0 \}
\prod_{i\in\mcal{I}}\bb{I}\{\tn_i\geq \tk_i \}
\bb{I}\{\sum_{i\in\mcal{I}}\tk_i=k_p\}.
\label{eq:prob2}
\end{multline}
When $I=2$, \cref{eq:prob2} reduces to a hypergeometric distribution 
\begin{multline}
p(\tk_i;k_p,\bm{\tn},\bm{\tk})=\frac{\binom{\tn_1}{\tk_1}\binom{n_p-\tn_1}{k_p-\tk_1}}{\binom{n_p}{k_p}}
\bb{I}\{\tk_1\geq 0, k_p- \tk_1\geq 0\}\\
\bb{I}\{\tn_1\geq \tk_1, n_p-\tn_1\geq k_p-\tk_1\}.
\end{multline}

\subsubsection{Utilities of EIPs}\label{sub:payoffs}
The utility or payoff of an EIP participating in the CEF is the difference between the rewards it receives and the costs it incurs in delivering the CDC services. 

\paragraph{Rewards}\label{para:reward}
Given a TP sends a type $p$ task to a nearby CEF, an EIP $i\in\mcal{I}$ that participates in this CEF can receive one or more of the following rewards. 
\begin{itemize} 
\item \textit{fixed reward} $R_i^{(2)}$, when an CEF is formed, based simply on EIP $i$'s participation in the CEF,
\item \textit{base reward} $R_i^{(0)}$, when the CEF satisfies the requirement of $n_p$ workers, based on the number $\tn_i$ of workers from EIP $i$ in executing the task,
\item and \textit{additional reward} $R_i^{(1)}$, when TP receives the $k_p$ sub-results, based on the number $\tk_i$ of EIP $i$ workers among the first $k_p$ workers to return the results to the TP.
\end{itemize}
With the reward scheme described above, we have
\begin{equation}
R^{(0)}_i = r_0 n_p \tn_i, \quad R^{(1)}_i = r_1 k_p \tk_i\text{, and} \quad R^{(2)}_i = r_2,
\end{equation}
where $r_0$, $r_1$, and $r_2$ are the unit reward for the base, additional, and fixed reward, respectively. Note that $R^{(0)}_i$ and $R^{(1)}_i$ are proportional to $n_p$ and $k_p$, due to the higher communication, storage, and computation costs incurred by an EIP when the values of $n_p$ and $k_p$ are higher.
%
%

\paragraph{Costs} \label{para:cost}
Given a TP sends type $p$ task to a nearby CEF and the edge cloud of EIP $i,i\in\mcal{I}$ that participates in this CEF contributes $l_i$ workers, the associated costs for the EIP are \textit{energy cost} $C_i^{(1),p}$ and \textit{edge resource utilization cost} $C_{i,l_i}^{(2)}$. First, energy cost to EIP $i$ when running a type $p$ task is proportional to the number $\tn_i$ of workers running the task,
\begin{equation}
C_i^{(1),p}=c_i \frac{D_p}{k_p}\tn_i,
\end{equation}
where $c_i$ is the cost per CPU cycle for EIP $i$. Second, edge resource utilization cost is to take into account the scarcity of the computing resources and thus amortize the fixed cost for an EIP, denoted by $C_i,i\in\mcal{I}$, to its different service provisions \cite{chandyAnalysisResourceCosts2009}, namely the CEF or the normal edge services in this paper. 
Let $W_i, i\in\mcal{I}$ denote the total edge capacity for EIP $i$, measured by the number of workers, and
$w_i$ denote the resource utilization, i.e., the fraction of EIP edge resources allocated to the CEF. Therefore, we have
\begin{equation}
w_i= \frac{E_i \sum_{l_i\in \mcal{L}_i} l_i x_{i,l_i}}{W_i}
\end{equation}
Here, we have assumed that edge clouds of EIP $i$ can simultaneously provide $L_i$ workers, i.e., $E_iL_i\leq W_i$. The resource utilization cost function \cite{chandyAnalysisResourceCosts2009} for EIP $i$ is 
\begin{equation}
f(w_i)= -C_i \rho_i \left( 1-\frac{1}{1-w_i}\right),
\end{equation}
where $\rho_i$ is a parameter between $0$ and $1$, representing the ratio of the fixed cost to be calibrated. The edge resource utilization cost for an edge cloud contribute $l_i$ workers to an CEF is defined as follows:
\begin{equation}
C_{i,l_i}^{(2)}=\frac{l_i x_{i,l_i}}{\sum_{j\in \mcal{L}_i}{jx_{i,j}}}\frac{f(w_i)}{E_i}
\end{equation}

\paragraph{Net Utility} \label{para:utility}
We first consider the homogeneous case when the tasks assigned to the CEF are all of the same type, e.g., $p$. Given an CEF formed among edge clouds of $I$ EIPs whose joint contribution of workers is $\bm{l}$, the expected utility that EIP $i$ can receive is
\begin{multline}
\pi_i^{(p)}(l_i;\bm{l})=R_i^{(2)} +
\bb{E}\left[ R_i^{(0)}-C^{(1),p}_i + \bb{E}[R^{(1)}_i|\tn_i]\right]  -C_{i,l_i}^{(2)}\\
= r_2
+ \sum_{\tn_i\in\mcal{L}_i}p(\tn_i;n_p,\bm{l},\bm{\tn}) \left(   r_0 n_p \tn_i
- c_i \frac{D_p}{k_p} \tn_i 
+ \sum_{0\leq \tk_i \leq \tn_i}  \right. \\
\left.
\sum_{\forall 0\leq \tk_j\leq \tn_j,j\neq i\in \mcal{I}}
p(\tk_i;k_p;\bm{\tn},\bm{\tk})r_1 k_p \tk_i \right) 
- \frac{l_i x_{i,l_i}}{\sum_{j\in \mcal{L}_i}{jx_{i,j}}}{f(w_i)}.
\label{homo-pure-payoff}    
\end{multline}

%
%
In the heterogeneous settings, when tasks assigned to the federation are of multiple types, i.e., $\{(n_p,k_p,D_p,\lambda_p)\}_{p\in\mcal{P}}$, the expected utility in \cref{homo-pure-payoff} is extended to
\begin{equation}
\pi_i(l_i;\bm{l})=\sum_{p\in\mathcal{P}} \frac{\lambda_p}{\sum_{q\in\mcal{P}} \lambda_q}  \pi_i^{(p)} (l_i;\bm{l}),
\label{payoff-ext}
\end{equation}
where $\frac{\lambda_p}{\sum_{q\in\mcal{P}} \lambda_q}$ indicates the probability that the type $p$ task is requested. 

Recall that the mixed strategy profile $\bm{x}$ defines the probability of having the pure-strategy profile $\bm{l}$. Let $\bm{l}_{-i}$ denote the pure strategy played by EIPs in $\mcal{I}$ other than $i$. 
Therefore, given EIP $i$ edge cloud contribute $l_i$ workers to an CEF, EIP $i$ expected payoff is
\begin{equation}
u_{i,l_i}(\bm{x})=\sum_{\forall l_j\in \mcal{L}_j, j\neq i\in \mcal{I}}  \pi_i(l_i;\bm{l})\prod_{j\neq i, j \in\mcal{I}}x_{j,l_j},
\label{eq:uij}
\end{equation}
where ${\forall l_j\in \mcal{L}_j,j\neq i\in \mcal{I}}$ enumerate all the instances for $\bm{l}_i$ given fixed $l_i$. When $I=2$, \cref{eq:uij} becomes
\begin{equation}
u_{i,l_i}(\bm{x})= \sum_{l_h\in\mathcal{L}_h} \pi_i(l_i;\bm{l}) x_{h,l_h},
\end{equation}
where $h$ is the other EIP in the game. 
%
%
%
Finally, the average payoff that EIP $i$ receives from an CEF is
\begin{equation}
u_i(\mathbf{x}) = \sum_{l_i\in\mcal{L}_{i}} u_{i,l_i}(\bm{x}) x_{i,l_i}.
\end{equation}

In summary, \cref{sub:system} presents the system model. In particular, we define the mixed strategy profile $\bm{x}$ and the payoff of each strategy for EIP $i$, i.e., $u_{i,j}(\mathbf{x}),j\in\mcal{L}_i,i\in\mcal{I}$. Both play key roles later in the formulation of the evolutionary game. Note that when there is no confusion, we replace $l_i$ by $j$ for convenience, e.g. $u_{i,j}(\mathbf{x}),j\in\mcal{L}_i$ is the same as that in \cref{eq:uij}. Moreover, \cref{eq:uij} indicates that the payoff for EIP $i$ is based on $\bm{x}$ and thus depends on all EIPs' decisions. The existence of an equilibrium point where no EIP can benefit by unilaterally altering its decision is therefore of interest. Next, we illustrate how to use game theory to identify such an equilibrium point.  

\subsection{Coded Edge Federation Formed as a Classical Evolutionary Game}\label{sub:game}
Game theory (GT) is a well-studied framework for analyzing the decision making process of various agents that have similar or conflicting objectives \cite{weibullEvolutionaryGameTheory1997}. The most common solution for a game is the Nash Equilibrium (NE), an equilibrium point at which no EIP can benefit by unilaterally changing its decision. To identify the NE, we adopt an evolutionary game approach due to its important property of bounded rationality. With the replicator dynamics, the dynamics behavior of EIPs can be represented by a system of differential equations, the solution of which is the equilibrium point in the game. In the following, we present the formulation of the evolutionary game (\cref{para:game-formulation}), an introduction of replicator dynamics (\cref{para:replicator-dynamics}), and the limitation of classical replicator dynamics (\cref{subsub:limitation}).


\subsubsection{Game Formulation}\label{para:game-formulation}
A game can be represented by a set of \textit{players}, a set of \textit{strategies} of each player, and the \textit{utility} of each strategy. In particular, they are the set of EIPs $\mcal{I}$, the set of strategies $\mcal{L}_i,\forall i\in \mcal{I}$, and the payoffs $u_{i,j}(\bm{x}),\forall i\in \mcal{I},j\in\mcal{L}_i$, where $\bm{x}$ is the mixed strategy profile, defined in the system model. 


In the context of the evolutionary game, edge clouds of EIP $i$ can be considered to be a \textit{population}. Thus, there are $I$ populations and the size of population $i$ is $E_i$. 
In addition, for an EIP $i$, its mixed strategy $\bm{x}$ is referred to as the \textit{population states} in evolutionary game \cite{weibullEvolutionaryGameTheory1997}. 
A strategy that leads to a higher payoff is more widely spread in the population, reflected by a large population states, after several rounds of strategic interactions among the EIPs. Therefore, if we randomly pick an edge cloud of EIP $i$, it is of higher chance, reflected by the mixed strategy, that the edge cloud adopts the successful strategy. In the following, we interchangeably use the terms of population states and mixed strategy.
%
Next, we present the revision protocol in the evolutionary game.

\subsubsection{Revision Protocols and the Classical Replicator Dynamics}\label{para:replicator-dynamics}
According to \cite{weibullEvolutionaryGameTheory1997}, the evolutionary game allows a strategy adjustment as follows: at each time instant, each cloud in the CEF has a revision opportunity to switch to some other strategy. 
Let $\rho^{(i)}_{j,k}$ denote the switch rate of a edge cloud of EIP $i$ in the CEF from strategy $j$ to $k$. Then the switch probability from $j$ to $k$ is proportional to $\rho^{(i)}_{j,k}$. The number of devices of EIP $i$ contributing $j$ workers can be modeled with the dynamics as follows:
\begin{equation}
\dot{x}_{i,j}= \gamma \sum_{k\in\mcal{L}_i}x_{i,k}\rho^{(i)}_{k,j}(\bm{x})-x_{i,j}\sum_{k\in\mcal{L}_i}\rho^{(i)}_{j,k}(\bm{x}),  \quad i\in\mcal{I},j\in\mcal{L}_i,
\label{markov}
\end{equation}
which also corresponds to the \textit{mean dynamics} \cite{sandholmPopulationGamesEvolutionary2010}. In \cref{markov}, $\dot{x}_{i,j}=\frac{d}{dt}x_{i,j}(t)$ is the first order time derivative. The first term in the right-hand side of \cref{markov} models the inflow of players switched from other strategies to strategy $j$, while the second term models the outflow of players from the strategy $j$ to the other strategies. Therefore, given a revision protocol $\rho$, \cref{markov} models the dynamics of a population involved in a strategic game. 

Although there are many types of protocols, a commonly used protocol is the \textit{proportional imitation protocol},  $\rho^{(i)}_{j,k}(\bm{x})=x_{i,k}[u_{i,k}(\bm{x})-u_{i,j}(\bm{x})]_+$, where $[x]_+=\max(0,x)$. In this protocol, when two random players with strategies $j$ and $k$ play against each other, a player with a lower payoff may learn from the other player by imitating his strategy, while the player with higher payoff does not change his strategy. The probability of imitation is proportional to the difference between the payoffs. Due to the wide application of this protocol \cite{niyatoDynamicsNetworkSelection2008,jiangGraphicalEvolutionaryGame2014,dongJointOptimizationTask2019c,cuiNovelMethodMobile2020}, we also adopt it in the classical evolutionary game for the CEF.
Together with \cref{markov}, the proportional imitation protocol leads to the so-called \textit{replicator dynamics}  
\begin{equation}
\dot{x}_{i,j}= \gamma x_{i,j}\left[u_{i,j}(\bm{x})-u_i(\bm{x}) \right],\quad i\in\mcal{I},j\in\mcal{L}_i,
	\label{replicator dynamics}
\end{equation}
where $x_{i,j}(t)\,\cond$ is the population state, $u_{i,j}(\bm{x})$ is defined in \cref{eq:uij}, and $\gamma$ is adaptation speed in the evolutionary process. 
With the replicator dynamics defined in \cref{replicator dynamics}, we can present the strategy adaptation process of an EIP as a system of ordinary differential equations, the solution to which is the stable mixed strategy profile of all EIPs in the CEF.

\subsubsection{Limitation of the Classical Replicator Dynamics}\label{subsub:limitation}
Although replicator dynamics are frequently used in engineering applications, it is myopic or memoryless  \cite{tarasovaConceptDynamicMemory2018}. In other words, in the CEF, when an EIP adjusts its mixed strategy at the current time, it forgets all the mixed strategies it used in the past. However, this assumption is not realistic, as the EIPs are economically rational market players who can use the information from the past to assist in future decision-making. 
%
The reason for the myopic feature is that integer-order derivatives are limit values, defined only in an infinitesimal neighborhood of a given time. Thus, the dynamics only capture the information in this small neighborhood and do not take into account the historical information before the present time \cite{weibullEvolutionaryGameTheory1997}. Therefore, the strategy adaptation of an EIP modeled by the replicator dynamics is memoryless, which restricts the application of the classical evolutionary game to the CEF in the real life. 

Next, we show how to overcome the memoryless limitation of classical replicator dynamics using fractional calculus.

%


\section{Fractional Evolutionary Game} \label{sec:fracgame}
%
To incorporate memory in the classical replicator dynamics, we use the left-sided Caputo derivatives, a particular type of fractional derivatives \cite{kilbasTheoryApplicationsFractional2006,podlubnyFractionalDifferentialEquations1998}, which allows past information to be aggregated in a logical way. We present the basics of fractional calculus in \cref{subsection: preliminarys -----}, the concept of power-law fading memory in \cref{sub:powerlaw}, and the formation of the fractional evolutionary game in \cref{sub:frac-game}.

\subsection{Basics of Fractional Calculus} \label{subsection: preliminarys -----}
We use the Caputo fractional derivative, since the Caputo fractional derivative of a constant is zero, and the initial conditions of the systems of Caputo fractional differential equations are integer-order derivatives \cite{kilbasTheoryApplicationsFractional2006}. These two features make the application of the Caputo derivative convenient, compared to other types of the fractional derivatives. However, the definition and the properties of the Caputo derivative still rely on the Riemann-Liouville fractional integrals and derivatives. For the convenience of the reader, we present the key concepts adapted from \cite{kilbasTheoryApplicationsFractional2006,podlubnyFractionalDifferentialEquations1998}. 
Let $\bb{N}=\{1,2,3,\ldots\}$ denote the set of positive integers, 
$\mathbb{N}_{0}=\{0\}\cup \mathbb{N}$ denote the set of non-negative integers, 
$[a,b]$ denote a closed interval of the real line $\bb{R}$, $AC^n[a,b]$ denote the space of real-valued functions $f(x)$ which have continuous derivatives up to order $n-1$ on $[a,b]$ such that $f^{(n-1)}(x)$ is absolutely continuous on $[a,b]$, and $[x]$ denote the largest integer that is smaller than or equal to $x$, e.g., $[3.1]=3$.

\begin{definition} [Riemann-Liouville Fractional Integrals] 
The \textit{(left-sided) Riemann-Liouville (RL) fractional integrals} $I^\alpha_{a+}$ of order $\alpha>0$ is defined as 
\begin{equation}
\left(I_{a+}^{\alpha} f\right)(x):=\frac{1}{\Gamma(\alpha)} \int_{a}^{x} \frac{f(t) d t}{(x-t)^{1-\alpha}}\quad (x>a ; \alpha>0).
\label{rl frac integral}
\end{equation}
In \cref{rl frac integral}, $\Gamma(z)=\int_{0}^{\infty} t^{z-1} e^{-t} d t \quad(\Re(z)>0)$ is the \textit{Euler Gamma function}. 
\label{def:rl integral}
\end{definition}

Note that for $\alpha\in\bb{N}$, \cref{rl frac integral} reduces to $n$th integrals of the form
\begin{multline}
\left(I_{a+}^{n} f\right)(x) =\int_{a}^{x} d t_{1} \int_{a}^{t_{1}} d t_{2} \cdots \int_{a}^{t_{n-1}} f\left(t_{n}\right) d t_{n} \\
=\frac{1}{(n-1) !} \int_{a}^{x}(x-t)^{n-1} f(t) d t \quad(n \in \mathbb{N}).
\end{multline}

\begin{definition}[Riemann-Liouville Fractional Derivatives] 
The \textit{Riemann-Liouville (RL) fractional derivatives} $D_{a+}^{\alpha}$ of order $\alpha>0$ is defined as 
\begin{multline}
\left(D_{a+}^{\alpha} y\right)(x):=\left(\frac{d}{d x}\right)^{n}\left(I_{a+}^{n-\alpha} y\right)(x)
=\frac{1}{\Gamma(n-\alpha)}\\
\left(\frac{d}{d x}\right)^{n} \int_{a}^{x} \frac{y(t) d t}{(x-t)^{\alpha-n+1}}\quad (n=[\alpha]+1 ; x>a),
\label{rl derivatives}
\end{multline}
where $I_{a+}^{n-\alpha}$ is the RL fractional integrals of order $n-\alpha$ given in \cref{rl frac integral}. 
Note that for $\alpha=n\in\bb{N}_0$, \cref{rl derivatives} reduces to the normal $n$th order derivatives, i.e., $\left(D_{a+}^{n} y\right)(x) = y^{(n)}(x) \quad (n\in\bb{N})$.
\label{def:rl derivatives}
\end{definition}

\begin{definition}[Caputo Fractional Derivatives]
Let $D_{a+}^{\alpha}[y(t)](x)\equiv\left(D_{a+}^{\alpha} y\right)(x)$ be the {RL fractional derivatives} of order $\alpha\geq 0$. Then the \textit{left-sided Caputo fractional derivative} of order $\alpha$ is defined as
\begin{equation}
\left({ }^{C} D_{a+}^{\alpha} y\right)(x):=\left(D_{a+}^{\alpha}\left[y(t)-\sum_{k=0}^{n-1} \frac{y^{(k)}(a)}{k !}(t-a)^{k}\right]\right)(x),
\label{caputo}
\end{equation}
where $n=[\alpha]+1$ for $\alpha \notin \mathbb{N}_{0}$ and $n=\alpha$ for $\alpha \in \mathbb{N}_{0}$. Note that for $\alpha=n\in\bb{N}_0$, the Caputo derivatives coincide with the normal integer-order derivatives, i.e., $({ }^{C} D_{a+}^{\alpha} y)(x) = y^{(n)}(x)$.
\label{def caputo}
\end{definition}

\begin{theorem}
If $y(x)\in AC^n[a,b]$, then the left-sided Caputo fractional derivatives ${ }^{C} D_{a+}^{\alpha}$ exist almost everywhere on $[a,b]$. In particular, for $\alpha \notin \bb{N}_0$,
\begin{multline}
\left({ }^{C} D_{a+}^{\alpha} y\right)(x) = \left(I_{a+}^{n-\alpha} D^{n} y\right)(x)\\
=\frac{1}{\Gamma(n-\alpha)} \int_{a}^{x} \frac{y^{(n)}(t) d t}{(x-t)^{\alpha-n+1}}, \label{eq: captuo def 2}  
\end{multline}
where $D=d/dx$ and $n=[\alpha]+1$. 
\label{captuo def 2}
\end{theorem}

Note that \cref{eq: captuo def 2} will be used in the theoretical analysis in \cref{section: equiblirum anlaysis} and \cref{caputo} in the experiments, due to its efficiency in implementation. A proof of \cref{captuo def 2} can be found in \cite{kilbasTheoryApplicationsFractional2006}.

\subsection{Power-law Fading Memory via Caputo Derivatives} \label{sub:powerlaw}
The memory effect has been intensively studied in economic processes \cite{teyssiereLongMemoryEconomics2006,tarasovaElasticityEconomicProcesses2016,tarasovaConceptDynamicMemory2018}, which comprise two types of time-dependent variables: \textit{exogenous} (independent) variables $X(t)$ and \textit{endogenous} (dependent) variables $Y(t)$. As endogenous variables describe the reaction to the exogenous variables, the most widely used dynamics to capture their relationship are the first-order time derivatives $Y(t)=X'(t)$. However, this expression only depends on the present time $t$ and is memoryless \cite{teyssiereLongMemoryEconomics2006}. 

To take into account the memory effect, the author in \cite{tarasovaConceptDynamicMemory2018} proposes the dynamic memory, represented by a general form of $Y(t)=F^t_0 (X'(\tau)),\,\forall \tau\in [0,t]$. Here, $F^t_0$ is an operator or a functional, which defines a certain method that identifies the value of $Y$ at time $t$ given the past information of $X'(\tau)$ for all $\tau\in[0,t]$. 
%
%
In particular, the author proposes to use the \textit{Volterra operator} on $X'(\tau)$, i.e., $F^t_0(X'(\tau)):=\int_0^tM(t-\tau)X'(\tau)d\tau$. In this way, the value of $Y(t)$ is an aggregated amount of the past information $X'(\tau),\tau\in[0,t]$ and depends not only the value of $X'(t)$ at the present time but also its previous states $X'(\tau),\tau\in[0,t)$. The kernel of the integral operator, i.e., the function $M(t-\tau)$, plays a key role. It is also called the \textit{memory function}, which represents the \textit{weight} of the historical information in the memory. 
It is worth noting that the weight $M(t-\tau)$ depends on the time distance between $\tau$ and $t$, and therefore it provides different weights to the different historical information in the memory. Besides $X'(t)$, the Volterra operator can be applied to the case of any integer-order derivatives $\frac{d^n}{d\tau^n}X(\tau)$ of $X(\tau)$ with respect to time $\tau$, i.e., $Y(t)=F^t_0(X^{(n)}(\tau))=\int_0^tM(t-\tau)X^{(n)}(\tau)d\tau \quad (n\in\mathbb{N}_0).$

To ensure that the dynamic memory is \textit{fading}, i.e., satisfying $M(t-\tau)\rightarrow 0 $ when $t\rightarrow +\infty$ with fixed $\tau$, the author in \cite{tarasovaConceptDynamicMemory2018} specifies the memory with \textit{power-law fading}, a concept that was first proposed by Volterra in \cite{daniellTheoryFunctionalsIntegral1932} and later played a key role in modern physics \cite{koellerApplicationsFractionalCalculus1984,colemanGeneralTheoryFading1968}. In addition, the power-law fading memory has been demonstrated in experiments intensively in \cite{edelmanEvolutionSystemsPowerLaw2020}, e.g., power-law human learning \cite{andersonLearningMemoryIntegrated2000} and power-law forgetting \cite{wixtedAnalyzingEmpiricalCourse1990}. The memory function with the power-law fading of order $\alpha$ is defined as \cite{tarasovaConceptDynamicMemory2018}
\begin{equation}
M(t-\tau)=M_{n-\alpha}(t-\tau)=\frac{1}{\Gamma(n-\alpha)}\frac{B}{(t-\tau)^{\alpha-n+1}},
\label{power-law-fading}
\end{equation}
where $n=[\alpha]+1$. With \cref{power-law-fading}, the economic process with \textit{power-law fading memory} function is $Y(t)= \left({ }^{C} D_{0+}^{\alpha} X \right)(t)$, where ${ }^{C} D_{0+}$ is the left-sided Caputo fractional derivative of the order $\alpha\geq 0$ defined in \cref{eq: captuo def 2}. 
Note that the Caputo fractional derivative of the order $\alpha=1$ coincides with the first order derivative (see \cref{def caputo}). Thus, the dynamics of power-law fading memory function generalize the standard dynamics of first-order derivatives to the fractional-order derivatives. Next, we formulate the evolutionary game based on the replicator dynamics that incorporate the power-law fading memory.

\subsection{Fractional Replicator Dynamics and the Fractional Evolutionary Game} \label{sub:frac-game}
To reflect the economic-aware nature of the EIPs, we incorporate the power-law fading memory in \cref{power-law-fading} to the replicator dynamics \cref{replicator dynamics} given by
\begin{multline}
	(\prescript{C}{}{{D}_{0+}^{\alpha}}x_{i,j})(t) =\gamma x_{i,j}(t)\left[u_{i,j}(\mathbf{x}(t))  -u_i(\mathbf{x}(t)) \right],\\ \quad i\in\mcal{I},j\in\mcal{L}_i, 0<\alpha<2.
	\label{fractional dynamics}
\end{multline}
The new dynamics is referred to as \textit{fractional replicator dynamics}, which depends not only on the population states at the present time $t$ but also their previous states in the evolutionary process. We refer to the evolutionary game with the fractional replicator dynamics in \cref{fractional dynamics} as \textit{fractional evolutionary game}. 
In particular, when $\alpha=1$, the fractional evolutionary game is equivalent to the classical evolutionary game; when $0<\alpha<1$, fractional replicator dynamics in \cref{fractional dynamics} capture the change of the strategies $\bm{x}$ in the evolutionary process, and is sometimes referred to as a \textit{subdiffusion} process; when $1<\alpha<2$, fractional replicator dynamics in \cref{fractional dynamics} capture the \textit{speed} of change of the strategies $\bm{x}$ in the evolutionary process. This is sometimes referred to as a \textit{superdiffusion} process. Next, we theoretically examine the validity of the fractional replicator dynamics.

\section{Equilibrium Analysis}
\label{section: equiblirum anlaysis}
Some properties of fractional replicator dynamics are examined in this section. We present the system of fractional ordinary equations in \cref{sec;system of fde}, the existence and uniqueness of the solution in \cref{sec;exist and unique}, and the evolutionary stability in \cref{sec;stablity}.

%
\subsection{The System of Fractional Differential Equations}\label{sec;system of fde}
Let mapping $\phi=
[\phi_{i,j}]_{i\in\mathcal{I},j\in\mcal{L}_i}$ with $\phi_{i,j}(x(t))=\gamma x_{i,j}\left[u_{i,j}(t)(\mathbf{x}(t))-u_i(\mathbf{x}(t)) \right]\, (t\in J)$ denote the right-hand side of the fractional replicator dynamics in \cref{fractional dynamics} and $J=[0,T]$ denote the time domain in the evolutionary process: 
%
Thus, \cref{fractional dynamics} can be represented by a system of fractional differential equations with initial values, also referred to as an initial value problem (IVP), defined by
\begin{align}
&{}^C D^{\alpha}_{0+} x_{i,j}(t) =\phi_{i,j}\left(\bm{x}(t)\right) \label{ivp-eq},\\
&x_{i,j}^{(k)}(0) =b_{i,j,k} \quad (\cond, k=0,1\ldots,[\alpha]+1), \label{ivp-value} 
\end{align}
where $b_{i,j,k}$ is the initial value of the corresponding $k$th integer-order differentiation for the population states $x_{i,j}$. With \cref{ivp-eq} and \cref{ivp-value}, the fractional evolutionary game becomes a system of fractional differential equations, the solution to which becomes the equilibrium points in the fractional evolutionary game.


The next lemma establishes an equivalence relation between the initial value problem and the integral problem. Further details can be found in   \cite{diethelmAnalysisFractionalDifferential2002a,daftardar-gejjiAnalysisSystemNonautonomous2007}.

\begin{lemma}
If the function $\phi_{i,j},\cond$ is continuous on its domain, then the initial value problem defined in \cref{ivp-eq,ivp-value} is equivalent to the nonlinear Volterra integral equation of the second kind,
\begin{multline}
x_{i,j}(t)=\sum_{k=0}^{n-1} \frac{t^{k}}{k !} x_{i,j}^{(k)}(0)+ \frac{1}{\Gamma(\alpha)} 
\int_{0}^{t}(t-s)^{
\alpha-1} \phi_{i,j}(\bm{x}(s)) ds \\ \quad (\cond,n=[\alpha]+1 ).
\label{voltera}
\end{multline}
\label{lemma equiv between volterra and fractional ode} 
\end{lemma}

In other words, every solution of the Volterra equation \cref{voltera} is also a solution of our original initial value problem defined by \cref{ivp-eq,ivp-value}, and vice versa. Thus, instead of solving initial value problem of \cref{ivp-eq,ivp-value}, we therefore focus on \cref{voltera}, which is more amenable to theoretical analysis. 
%
Note that \cref{voltera} is weakly singular if $0< \alpha < 1$ and regular for $\alpha > 1$. We focus on $0<\alpha<1$ in the following; the proof for $1<\alpha<2$ is similar.

\subsection{Existence and Uniqueness of the Solution}\label{sec;exist and unique}
To prove the existence and uniqueness of the solution to the fractional evolutionary game, we construct a complete metric space with $\norm{\cdot }^*$ by Lemma \ref{lemma equiv between volterra and fractional ode} and then prove the key result in \cref{main theorem}.

The solution trajectory $x_{i,j}(t)\,(\cond)$ is easily verified as a continuous mapping defined on the domain $J$. Let $C(J,\bb{R}^L)$ be the class of all continuous column vector functions $\bm{x}(t)=[x_{i,j}(t)]_{\cond}$ defined on $J$. 
Let $\|\bm{x}\|^{*}=\sum_{\cond} \sup _{t} e^{-N t}\left|x_{i,j}(t)\right|\, (N>0)$  be a new norm on $C(J,\bb{R}^L)$ and the associated metric is defined by $d(\bm{x}_1,\bm{x}_2)=\norm{\bm{x}_1-\bm{x}_2}^*$. 

\begin{lemma}\label{norm}
The norm $\norm{\cdot}^*$ is equivalent to the supremum norm on the space $C(J,\bb{R}^L)$.
\end{lemma}
The proof is given in Appendix \ref{app:norm}. Lemma \ref{norm} shows that the function space $\left\langle C(J,\bb{R}^L),d\right\rangle $ is a \textit{metric complete space}. Note that the constructed equivalent norms and the respective metrics are the main tools in the proof of existence and uniqueness for the fractional evolutionary game. 
The method of equivalent norms is widely used in the theory of differential equations 
\cite{baleanuGlobalExistenceSolutions2010,el-raheemModificationApplicationContraction2003}. 
Next, we show that the mapping $\bm{x}$ is a contraction on $\left\langle C(J,\bb{R}^L),d\right\rangle $ in \cref{lip corollary} and \cref{main theorem}.


\begin{theorem} The right-hand functions of the fractional replicator dynamics $\phi$  satisfy the Lipschizian conditions, i.e., for any $\bm{x},\bm{y} \in \C$,
\begin{multline}
\left|\phi_{i,j}\left(\bm{x}(t)\right)-\phi_{i,j}\left(\bm{y}(t)\right)\right|
\leq \sum_{\cond} A_{i,j}\left|x_{i,j}(t)-y_{i,j}(t)\right| \\
\quad A_{i,j}>0,\cond
\end{multline}
\label{lip corollary}
\vspace{-0.5cm}
\end{theorem}
The proof is given in Appendix \ref{appendix lipschiz}. With \cref{lip corollary}, we can obtain the key result that there exists a unique equilibrium point in the fractional evolutionary game as follows.

\begin{theorem}
With \cref{lip corollary}, \cref{voltera} has a unique solution $\bm{x}\in C(J,\bb{R}^L)$.
\label{main theorem}
\end{theorem}
The proof is given in Appendix \ref{appendix unique}.

\subsection{Stability of the Solution}\label{sec;stablity}
After verifying the existence and uniqueness of the equilibrium point in the fractional evolutionary game, we investigate the stability of the solution. 

\begin{definition}[Uniformly stable (See \cite{abdel-salamStabilityFractionalorderNonautonomous2007})] 
The solution to the fractional evolutionary game defined by \cref{ivp-eq,ivp-value} is said to be \textit{uniformly stable} if for any $\epsilon >0$ there exists $ \delta(\epsilon)>0$ such that for every initial point $\bm{y}_o\in\bb{R}^L$ satisfying $\norm{\bx_o-\bm{y}_o}_L<\delta$, we have 
$\norm{\bx(t)-\bm{y}(t)}^*<\epsilon$. 
Here, $\bm{y}(t)$ is the solution of initial value problem defined by \cref{ivp-eq} and the initial value given by \cref{new initial value}
\begin{equation}
\bm{y}_o=[\tilde{b}_{i,j,1}]_{\cond}.
\label{new initial value}
\vspace{-0.5cm}
\end{equation}
\label{def stabilty}
\end{definition}
\begin{theorem}
The solution of the fractional evolutionary game given by \cref{ivp-eq,ivp-value} is uniformly stable.
\label{key theorem stable}
\end{theorem}

The proof is given in Appendix \ref{appendix stability}. 

Next, we verify the above theoretical findings as well as the dynamic behaviors of the EIPs in the CEF through experiments.


\begin{table}
\caption{Parameter Values for EIPs}
\begin{center}
\begin{tabular}{ l c c c c c c c c c }
 EIP $i$ & $E_i$ & $L_i$ & $C_i$ & $\rho_i$ & $c_i$ & $W_i$\\ 
\hline
1 & 100 & 4  & 1800 & 1 & $10^{-5}$ & 500\\  
2 & 120 & 8  & 2800 & 1 & $10^{-5}$ & 1100\\  
\end{tabular}
\end{center}
\label{tab: eip parameter}
\end{table}

\begin{table}[]
\caption{Parameter Values for the Task Publishers}
\begin{center}
\begin{tabular}{ l c c c c c c c c c }
 task & id & $n$ & $k$ & $r_0$ & $r_1$ & $r_2$ & $D$ & $\lambda$ \\ 
\hline
homogeneous  & 1 & 6 & 4 & 30 & 30 & 10 & $10^{6}$  & 1\\  
\end{tabular}
\end{center}
\label{tab: hetero task paramter}
\end{table}

\section{Performance Evaluation}
\label{sec: performance evalutation}
In this section, we use numerical simulations to investigate the dynamic behavior of EIPs in the CEF. We consider the fractional replicator dynamics of three types, including the classical replicator dynamics ($\alpha=1$), superdiffusion processes ($\alpha=[1.2,1.4]$), and subdiffusion processes ($\alpha=[0.65,0.8]$). Unless otherwise stated, the simulated system consists of $2$ EIPs, i.e., EIP $1$ and EIP $2$, with parameter values as listed in \cref{tab: eip parameter}.
%
%
%
The domain of the trajectory is $[0,T]$, where $T=1$, and the maximum number of time steps is $10^4$. 
Table~\ref{tab: hetero task paramter} summarizes the task parameters and we omit the subscript $p$ due to the homogeneous tasks assumed. In the following, we investigate the properties of the power-law fading memory function (\cref{sec:impact of the memory weight}) and equilibrium existence and convergence rate (\cref{sec:existence of equilblrim}). Subsequently, we examine the stability of the equilibrium point (\cref{sec: exp-stability}) and the adaptation of the equilibrium when certain hyper-parameter values are varied (\cref{sec:sensititive}).


\subsection{The Memory Weight in Caputo Derivatives}
\label{sec:impact of the memory weight} 

\begin{figure}
	\begin{subfigure}[b]{0.49\linewidth}
		\includegraphics[width=\textwidth]{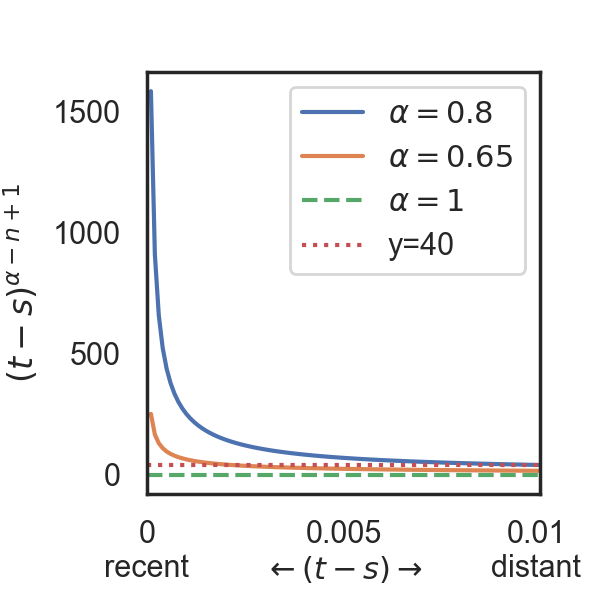}
		\caption{Memory weight for short-term memory}
		\label{fig:memory1}
	\end{subfigure}
	\hfill
	\begin{subfigure}[b]{0.485\linewidth}
		\includegraphics[width=\textwidth]{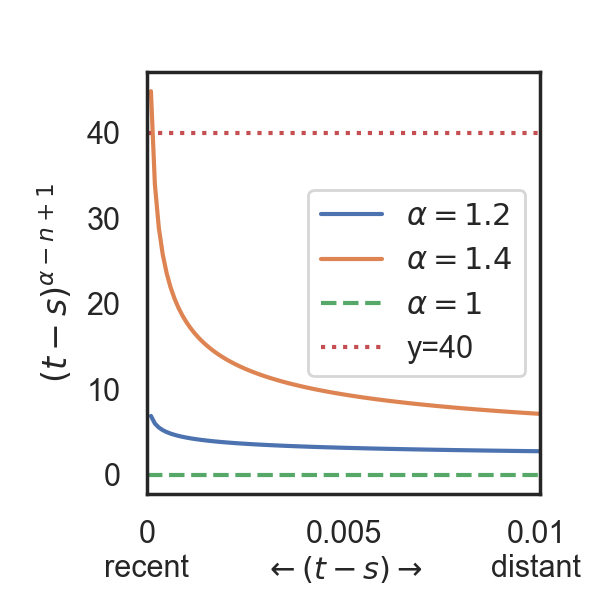}
		\caption{Memory weight for long-term memory}
		\label{fig:memory2}
	\end{subfigure}
	\caption{Power-law fading memory function vs. the time difference $(t-s)$}
	\label{figmemory}
\end{figure}
\begin{figure*}
\begin{minipage}{0.34\textwidth}

\end{minipage}
\hfill
\begin{minipage}{0.32\textwidth}
\centering
	\includegraphics[width=\linewidth]{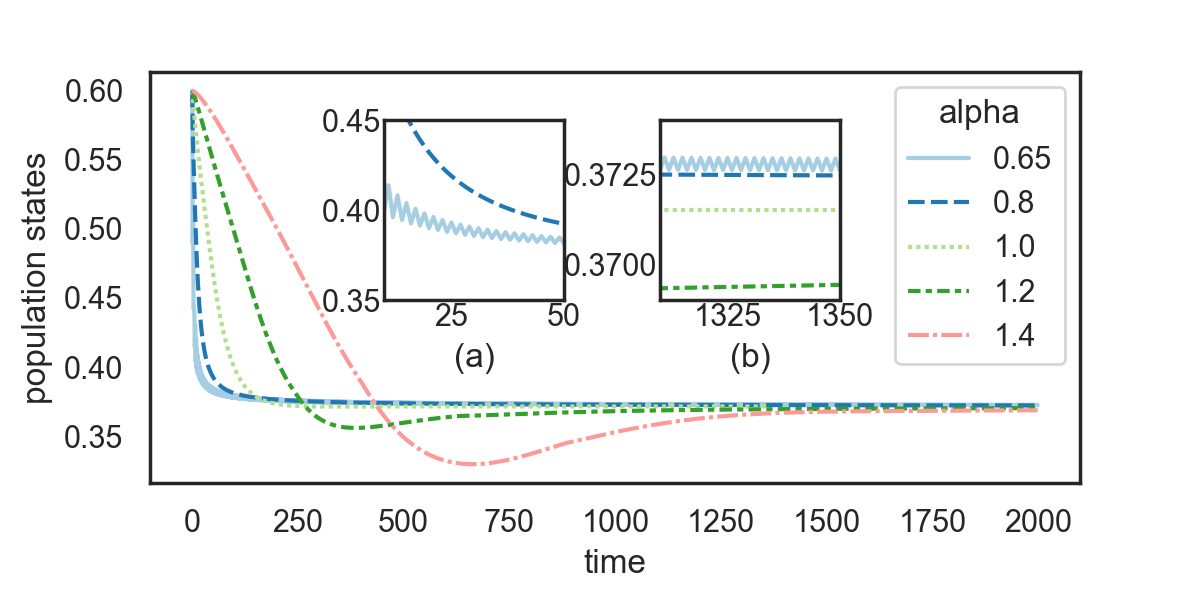}
	\caption{The trajectory of the last strategy of EIP $1$ in different types of the evolutionary process. Here, the code configuration is $n=6,k=4$.}
	\label{figevolve}
\end{minipage}
\hfill
\begin{minipage}{0.32\textwidth}
\centering
	\includegraphics[width=\linewidth]{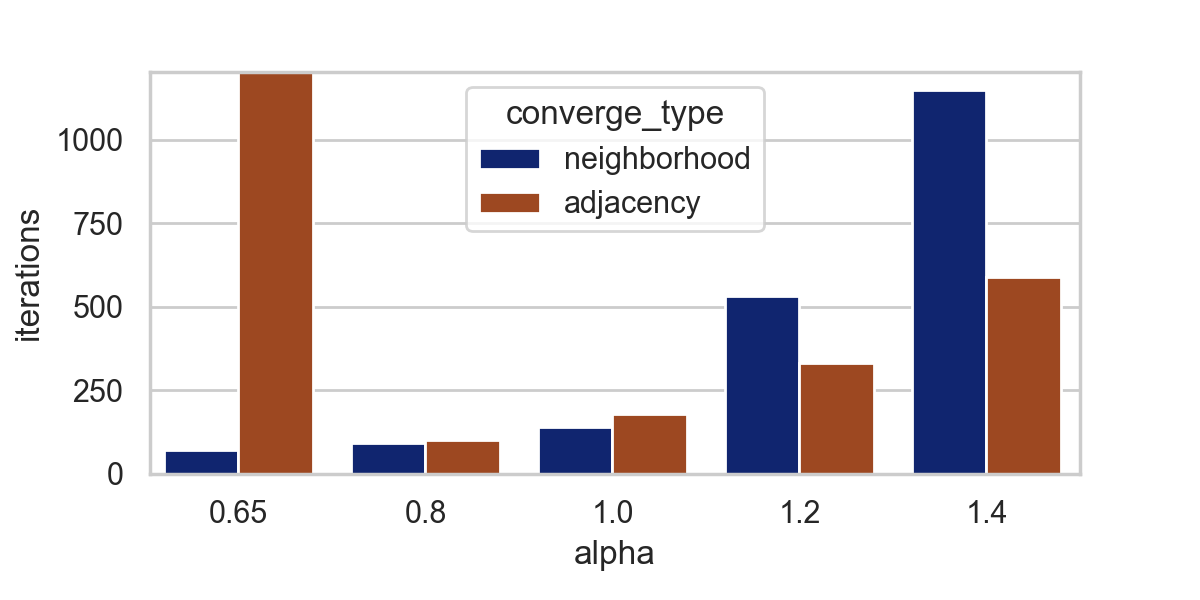}
	\caption{Time to converge of two types for different evolutionary processes. Here, the code configuration is $n=6,k=4$.}
	\label{figconverge}
\end{minipage}
\begin{minipage}{0.32\textwidth}
\centering
	\includegraphics[width=\linewidth]{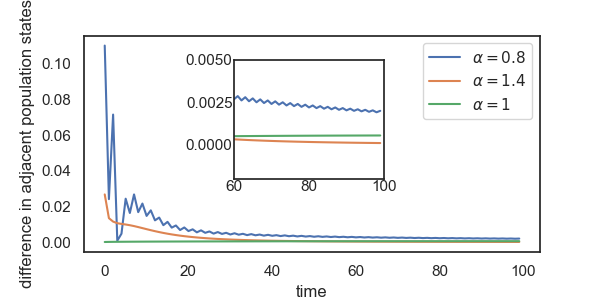}
	\caption{Change of the last strategy used by EIP $1$ as time evolves.} 
	\label{figdiff}
\end{minipage}
\end{figure*}

The memory weight, $(t-s)^{\alpha-n+1}$, utilized in the Caputo derivatives in \eq{eq: captuo def 2} is plotted against the time lag, $t-s$, in \cref{figmemory}. Here, $t$ represents the current timestamp, and $s$ is some historical timestamp before $t$. Thus, $t-s$ in fact represents how distant the historical information is. The larger the value of $t-s$ is, the more distant the historical information is. In addition, in \cref{figmemory}, we draw 
the memory function of the classical evolutionary game ($\alpha=1$) and a horizontal line with memory weight being $40$ as two reference lines.  

It can be observed from \cref{figmemory} that for both superdiffusion $(\alpha>1)$ and subdiffusion processes $(\alpha<1)$, a certain weight is placed on the memory about the historical information, since $(t-s)^{\alpha-n+1}$ is always positive. 
Additionally, \cref{fig:memory1} shows that in the subdiffusion processes ($\alpha\in [0.65,0.8]$) 
the memory weight decreases rapidly with time lag.
This indicates that the memory is in fact \textit{short-term}, since the very recent memory dominates the aggregated memory.
In contrast, \cref{fig:memory2} shows that in the superdiffusion process ($\alpha\in[1.2,1.4]$), the memory is rather long. 
For instance, the memory function of $\alpha=1.2$ in \cref{fig:memory2} is nearly flat, and thus there is little discrimination between the distant and recent historical information. Therefore, the memory, i.e., to aggregate the historical information, is long. 
For $\alpha=1.4$, the weight discrimination between the recent and past information is not significant either, compared to those in \cref{fig:memory1}. The range of the memory weight is around $[0,40]$ in \cref{fig:memory2} for $\alpha\in[1.2,1.4]$, compared to $[0,1500]$ for $\alpha\in[0.65,0.8]$ in \cref{fig:memory1}. 
Thus, with $\alpha\in[0.65,0.8,1,1.2,1.4]$, fractional replicator dynamics are short-term memory-aware in subdiffusion processes and long-term memory-dependent in the superdiffusion process in the CEF game.

Note that the value of $(t-s)^{\alpha-n+1}$ is the same for both $\alpha=0.6$ and $\alpha=1.6$ since $n=[\alpha]+1$. However, based on the definition of the Caputo derivatives in \eq{eq: captuo def 2}, the contents of the memory differ. 
In the subdiffusion process, it is the \textit{change} of past decisions, $y^{(1)}(t)$, to be stored in the memory kernel, while the \textit{rate of change} or the \textit{acceleration speed}, $y^{(2)}(t)$, is to be memorized in the superdiffusion process. Hence, even though the memory weights are the same for the values of $\alpha=0.6$ and $\alpha=1.6$, their evolutionary patterns can be very different. 

In conclusion, $\alpha$ determines what information is stored in the memory and the weight of recent past information relative to the distant one in the memory.

\subsection{Existence of Equilibrium and the Convergence Rate}
\label{sec: exp-equi-exist}

We now analyze the equilibrium point in terms of its existence and uniqueness for different values of $\alpha$. In particular, we consider the case where homogeneous tasks assigned to the CEF. The task parameter values are shown in \cref{tab: hetero task paramter}. 
First, we review and clarify the concept of equilibrium and convergence of two types in \cref{sec:type of convergence}. Second, we illustrate the existence and uniqueness of equilibrium by plots of trajectories in \cref{sec:existence of equilblrim}. Finally, we analyze the convergence pattern for different values of $\alpha$ in \cref{sec:pattern of convergence}.

\subsubsection{Equilibrium and Convergence}
\label{sec:type of convergence}
To better describe how processes reach the equilibrium points, we introduce the following two definitions: 
\begin{itemize}
	\item \textit{Adjacency Type of Convergence}: This concept is the same as the condition where the equilibrium point is reached. It gives the timestamp after which the difference between \textit{adjacent population states} is within some given threshold. We define the adjacency threshold as $0.0001$. Unless otherwise stated, we simply use \textit{convergence} to refer to this type in our paper.
	\item \textit{Neighborhood Type of Convergence}: This concept gives the timestamp after which the population states fall into the \textit{equilibrium neighborhood}, i.e., some points within a certain threshold of the equilibrium point. We set the neighborhood threshold at $0.01$. 
\end{itemize} 
The convergence describes an evolutionary process reaching the equilibrium in an exact sense, while the neighborhood type can provide an additional aspect to understand the pattern of convergence in a loose manner. 

\subsubsection{Existence of Equilibrium}
\label{sec:existence of equilblrim}

To examine the existence of the equilibrium, we plot the evolutionary trajectories of the last strategy used by EIP $1$ in \cref{figevolve} for all evolutionary processes with $\alpha\in\{0.65,0.8,1,1.2,1.4\}$. The last strategy is to contribute $4$ workers, which is the maximum number of workers for EIP $1$, from its edge clouds to the CEF. As shown in \cref{figevolve}, the algorithm of all evolutionary processes stops before reaching the maximum number of iterations, and the trajectory curves converge to a point around $0.37$. This illustrates the existence and uniqueness of the equilibrium in the fractional evolutionary game.

\subsubsection{Impact of $\alpha$ on the Rate of Convergence}
\label{sec:pattern of convergence}
Besides the existence of equilibrium, \cref{figevolve} shows different convergence rates for different evolutionary processes. For example, the process with $\alpha=0.65$ seems to converge fastest to the equilibrium neighborhood, and the process with $\alpha=1.4$ appears to converge slowest and fluctuates most. 

To better understand how different processes converge, we plot the time to converge for both the neighborhood type and adjacency type in \fig{figconverge}. The figure shows that the time to converge of the \textit{neighborhood} type decreases as $\alpha$ decreases.
In other words, it takes longer for the superdiffusion process to reach the \textit{equilibrium neighborhood} and shorter for the subdiffusion process compared to the classical evolutionary game. A similar pattern can be observed for the adjacency type of convergence, except for the case with $\alpha=0.65$. Particularly, the convergence for the subdiffusion process with $\alpha=0.65$ is dramatically longer, though it reaches the equilibrium neighborhood in the shortest time. Detailed explanations are provided in \sect{para:decrease} and \sect{para:minimum alpha}.

\paragraph{Faster Convergence as Value of $\alpha$ Decreases} \label{para:decrease}
Recall that an equilibrium is defined as certain population states, if the trajectories start with them are stationary. In other words, there is no change in decisions if both the EIPs' starting strategies are at equilibrium points in the CEF game. 

For the superdiffusion process, the replicators memorize the rate of change of past information. Thus, to reach the equilibrium, the trajectory curves must \textit{oscillate} around $0$ with a decreasing range. 
In other words, the rate of change, i.e., the acceleration velocity, fluctuates around $0$ with a decreasing range. Otherwise, the acceleration velocity is either continuously positive or continuously negative. 
As a result, the trajectory curve is either convex upwards or concave downward. In neither case, the trajectory can be stationary, indicated by a horizontal line. Due to this fluctuation nature of the superdiffusion process, a process where the impact from the recent past information is reduced, can help speed up the convergence. This effect happens when a longer memory of the past exists, represented by the decreasing value of $\alpha$. 

The above abstract analysis is consistent with the experiment findings as shown in \cref{figconverge} and \cref{figevolve}, i.e., the superprocess with $\alpha=1.2$ converges faster than the one with $\alpha=1.4$. This also has real-life implications when strategy adjustments for the EIP are driven by the rate of change of the information in the past ($\alpha>1$). Such EIPs are sensitive to the strategy adaptation and therefore, are harder to stabilize their strategy in a shorter time. 
However, if the EIP average past oscillation information over a longer time span, indicated by a smaller $\alpha$ in $(1,2)$, the impact from the recent oscillation is reduced, and the convergence velocity can be much improved.


%



\begin{figure}
\centering
\begin{minipage}{0.5\textwidth}

\end{minipage}
\vspace{-0.1cm}
\end{figure}

As for the subdiffusion process with $\alpha\in [0.65,0.8]$, we plot the change in population state in \cref{figdiff}. Unlike the superdiffusion process, the rate of change for the subdiffusion process can be continuously positive, continuously negative (e.g., $\alpha=0.8$ in \cref{figdiff}), and fluctuated ($\alpha=0.8$ in \cref{figdiff}). Additionally, in comparison to the superdiffusion process ($\alpha=1.4$), the early changes in population states in the subdiffusion process are more significant. 
This is due to the amplification of the recent past information in the memory kernel as introduced in \cref{sec:impact of the memory weight}. The lower the value of $\alpha$ is, the longer range of the amplified past information. Thus, with the recent past information, the EIP quickly identify the right direction to adjust its strategy, which accelerates the convergence of the evolutionary process.

\paragraph{The Minimum Value of $\alpha$}\label{para:minimum alpha} 
Although smaller values of $\alpha$ can lead to faster convergence for both superdiffusion and subdiffusion processes, a small $\alpha$ value may over-amplify the recent past information, causing the EIP to over-adjust the strategy in the early stage. Consequently, the EIP needs to reverse some of its adjustments in the next step, which is represented as fluctuating population states in the early stage. Given the longer memory effect from the smaller value of $\alpha$, the impact of such fluctuation lasts longer. As a result, more iterations are required to dilute the impact of the oscillation to reach equilibrium. In conclusion, there is some value of $\alpha$, below which the subdiffusion process start to show the a pattern of fluctuations. 

This also explains the oscillation curve for $\alpha=0.65$ and the smooth curve for $\alpha=0.8$ in \cref{figevolve} and \cref{figdiff}. Specifically, for $\alpha=0.65$, the evolution of the strategy for EIP $1$ demonstrates an oscillating pattern with a small range. With more iterations, the range of the fluctuation decreases, and the process eventually converges. The small fluctuations increase the convergence time.  

Our experiments demonstrate that $0.65$ is already near the minimum point at which such oscillations begin to appear. We observe fluctuations with a broader range for $\alpha=0.5$, and the trajectory does not converge within $10^4$ iterations. A longer iterative process is needed, which is beyond the settings of our experiments. Unless otherwise stated, we focus on the values of $\alpha$ from $0.7$ to $1.4$.

To summarize \cref{sec:pattern of convergence}, i.e., the impact of $\alpha$ to the pattern of convergence, we can conclude that a smaller value of $\alpha$ can accelerate the evolutionary process of the EIP's strategy adaptation. However, the value of $\alpha$ should not be too small for the subdiffusion process. Otherwise, the fluctuations around the equilibrium make the process take longer or even fail to converge in an exact sense.

\subsection{Evolutionary Stability of the Equilibrium}
\label{sec: exp-stability}
\begin{figure}
\begin{minipage}{0.5\textwidth}
	\centering
	\begin{subfigure}[b]{0.3\textwidth}
		\includegraphics[width=\textwidth]{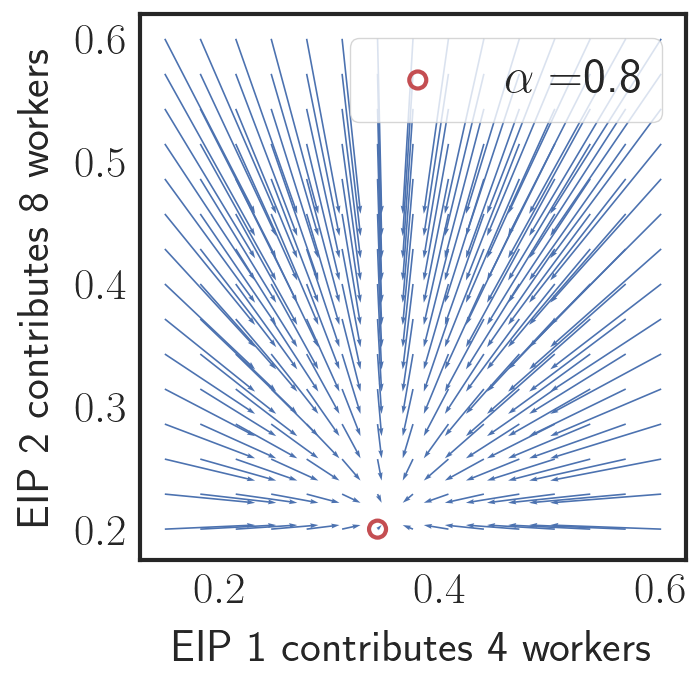}
		\caption{}
		\label{fig:41}
	\end{subfigure}
	\begin{subfigure}[b]{0.3\textwidth}
		\includegraphics[width=\textwidth]{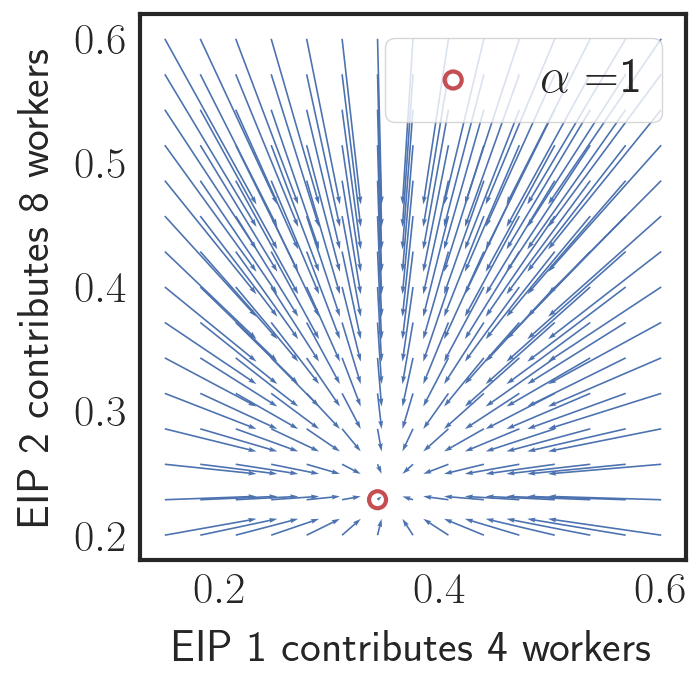}
		\caption{}
		\label{fig:42}
	\end{subfigure}
	\begin{subfigure}[b]{0.3\textwidth}
		\includegraphics[width=\textwidth]{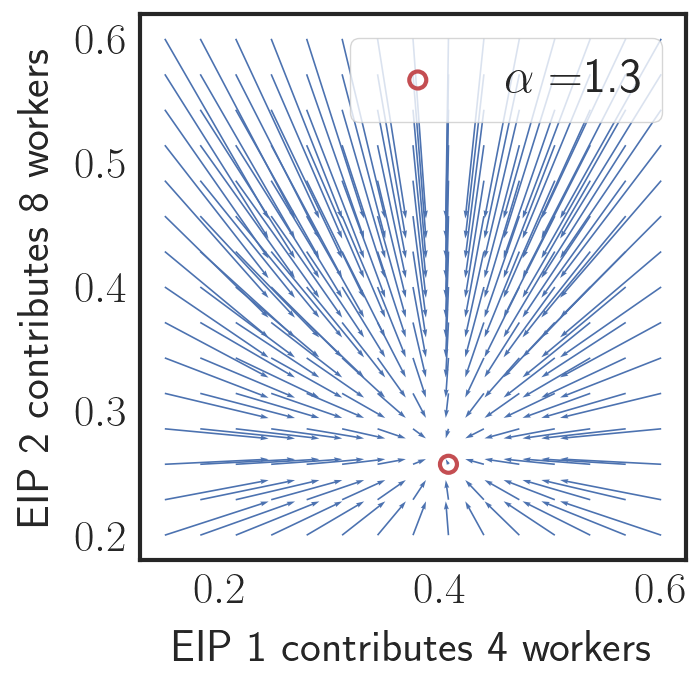}
		\caption{}
		\label{fig:43}
	\end{subfigure}
	\caption{The direction of evolutionary processes with different values of $\alpha$ and initial strategies.}
	\label{fig:stability} 
\end{minipage}
\vspace{-0.1cm}
\end{figure}
Next, we examine the evolutionary stability of the equilibrium point in the fractional evolutionary game for different values of $\alpha$. Fig.~\ref{fig:stability} shows the direction fields of the fractional replicator dynamics in steps of $200$ iterations for different initial population states. We select $\alpha=0.8,1,1.3$ to represent the subdiffusion process, classical evolutionary game, and the superdiffusion process, respectively. In addition, we vary the initial states of the last strategy for each EIP from $0.2$ to $0.6$ and keep the other strategies unchanged.

As shown in \cref{fig:stability}, all equilibrium points obtained in the fractional evolutionarily game, marked by the red circle, are evolutionarily stable. In particular, strategies that are not in equilibrium can follow the arrows indicated by the direction field and eventually reach equilibrium points. 
Strategies in equilibrium are thus robust to mutation and deviation within a certain threshold. This is consistent with the convergence and stability of the equilibrium point derived in \cref{key theorem stable}.

\begin{figure*}[]
\begin{minipage}{0.24\textwidth}
\centering
\begin{subfigure}[b]{\textwidth}
	\centering
	\includegraphics[width=\linewidth]{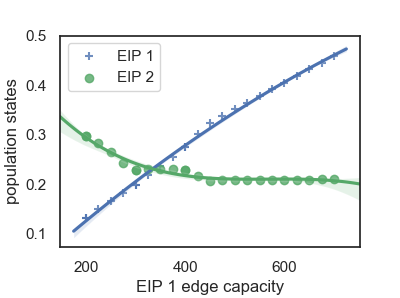}
	\caption{population states of the last strategy}
	\label{figcap1a}
\end{subfigure}

\begin{subfigure}[b]{\textwidth}
	\centering
	\includegraphics[width=\linewidth]{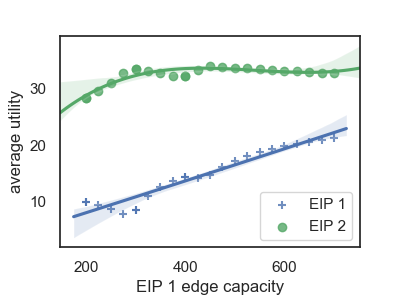}
	\caption{average utility}
	\label{figcap1b}
\end{subfigure}
	\caption{Impact of edge capacity.}
	\label{figcap}
\end{minipage}
\hfill
\begin{minipage}{0.24\textwidth}
\centering
	\begin{subfigure}[b]{\textwidth}
		\centering
		\includegraphics[width=\linewidth]{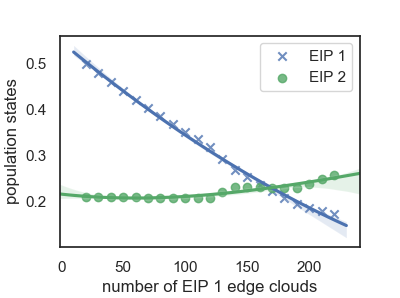}
		\caption{population states of the last strategy}
		\label{figns1a}
	\end{subfigure}

	\begin{subfigure}[b]{\textwidth}
		\centering
		\includegraphics[width=\linewidth]{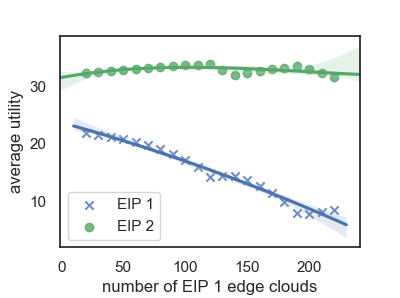}
		\caption{average utility}
		\label{figns1b}
	\end{subfigure}
	\caption{Impact of the number of edge clouds.}
	\label{figns}
\end{minipage}
\hfill
\begin{minipage}{0.24\textwidth}
\centering
	\begin{subfigure}[b]{\textwidth}
		\centering
		\includegraphics[width=\linewidth]{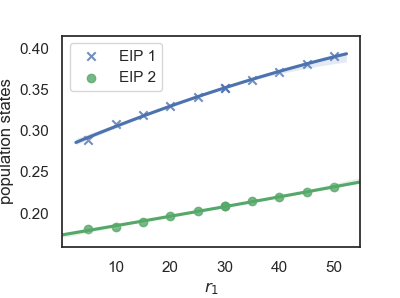}
		\caption{population states of the last strategy}
		\label{figr1}
	\end{subfigure}

	\begin{subfigure}[b]{\textwidth}
		\centering
		\includegraphics[width=\linewidth]{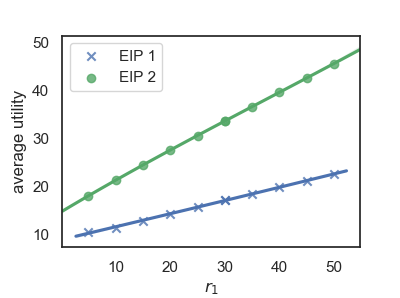}
		\caption{average utility}
		\label{figr2}
	\end{subfigure}
	\caption{Impact of unit of reward.}
	\label{figr}
\end{minipage}
\hfill
\begin{minipage}{0.25\textwidth}
\centering
	\begin{subfigure}[b]{\textwidth}
		\centering
		\includegraphics[width=\linewidth]{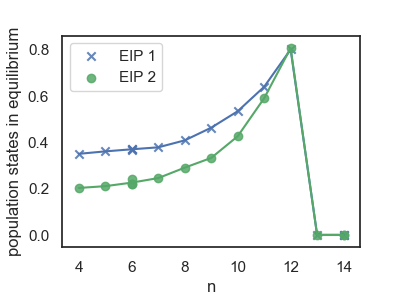}
		\caption{population states of the last strategy with $k=4$ and $n\in[4,14]$}
		\label{fign1}
	\end{subfigure}
\centering
	\begin{subfigure}[b]{\textwidth}
		\centering
		\includegraphics[width=\linewidth]{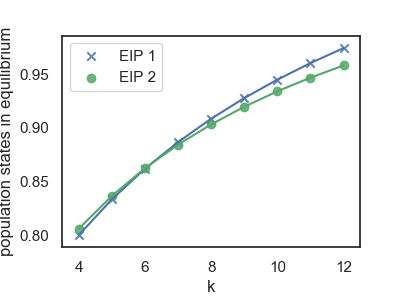}
		\caption{population states of the last strategy with $n=12$ and $k\in[4,12]$}
		\label{fign2}
	\end{subfigure}
	\caption{Impact of code configuration.}
	\label{fign}
\end{minipage}
\end{figure*}

\subsection{Equilibrium Adaptation}\label{sec:sensititive}
Having verified the properties of equilibrium in terms of its existence, uniqueness, and stability, we now examine the equilibrium adaptation when certain system parameter values are changed. In particular, we are interested in the edge capacity of each EIP (\sect{exp:imp_of_edge_capcity}), the number of edge clouds of each EIP (\sect{exp:imp_of_numer_of_edge_devices}), the reward (\sect{exp:imp_of_reward}), and the code configuration of the task publisher (\sect{exp:imp_of_code}).

\subsubsection{Impact of the Edge Capacity}
\label{exp:imp_of_edge_capcity}
Recall that edge capacity is defined as the EIP's maximum total number of workers that it can contribute to the CEF. Here, we vary the edge capacity of EIP $1$ while keeping that of EIP $2$ fixed. 

Figure \ref{figcap}\subref{figcap1a} shows that when EIP $1$ edge capacity increases, the probability of contributing $4$ workers from its edge clouds increases. The reason for this is that as EIP $1$ has more resources, it is more likely to fulfill the TP's requirements and earn a reward.
On the other hand, we can see that the population states of the last strategy of EIP $2$ decline because given the same task requirement, EIP $2$ can contribute fewer workers, but overall the federation can still meet the requirements of the TP and thus earn a payoff. 

Figure \ref{figcap}\subref{figcap1b} shows the average utility received by each EIP as the edge capacity of EIP $1$ increases. 
We observe that the average utility of each EIP increases if the edge capacity of EIP $1$ increases. 
Interpreting the upward trend for the average utility of EIP $1$ is straightforward. Given a larger probability of adopting the last strategy, workers of EIP $1$ on average have a greater presence in the federation and hence have a greater chance of being selected by the TP. 
As a result, the average utility received by EIP $1$ increases. On the other hand, the increasing contribution of EIP $1$ also benefits EIP $2$, as EIP $2$ can have a similar chance of meeting the TP' s requirement with fewer workers and correspondingly lower the cost for EIP $2$. 

In summary, \cref{figcap} reveals several benefits of the CEF. First, when one party, EIP $1$, increases its edge resource contribution, the other party, EIP $2$, can also enjoy a benefit, as reflected by a higher average utility. Second, the reward allocation scheme is fair, given that EIP $1$ enjoys a larger increase in the payoff relative to EIP $2$. Third, the scheme deters EIP $2$ from being a free-rider, as its amount of free benefit decreases as the share of EIP $1$'s workers in the CEF increases. The scheme thus incentivizes EIP $2$ to increase its edge resources contribution to maintain its market share in the CEF, thereby helping the CEF ecosystem to flourish. 
All the benefits help the federation draw and sustain EIPs to participate, incentivize EIPs to scale up edge resources investment, and ultimately drive the CEF ecosystem.

\subsubsection{Impact of the Number of Edge Clouds}
\label{exp:imp_of_numer_of_edge_devices}
This section examines the impact of the number of geo-diversified edge clouds in each EIP, given the edge capacity is fixed. With more geo-diversified clouds spread at the edge, the EIP coverage area becomes larger and receives more service requests from the TPs. In particular, we fix the number of edge clouds of EIP $2$ and vary the number of edge clouds of EIP $1$. 

Figure \ref{figns}\subref{figns1a} shows that as the number of edge clouds of EIP $1$ increases, the percentage of edge clouds configured to contribute $4$ workers is dramatically decreased. This is due to the fact that more services requests sent to the EIPs incurs higher edge resources utilization costs, i.e., a higher cost of allocating the scarce resources to the CEF. Therefore, from \cref{figns1b}, we see the average utility of joining the CEF decreases for EIP $1$ as the number of edge clouds increases. 
%
Meanwhile, \cref{figns1a} shows that while EIP $1$ reduces its probability of contributing $4$ workers in its mixed strategy (reflected by the population states) to the federation, EIP $2$ increases the probability of investing $8$ workers in the federation. The reason is that investing more workers can help EIP $2$ improve the chance of fulfilling the requirement of the TP and receiving the reward. 
Thus, we see a trend of increasing population state for EIP $2$. 

\subsubsection{Impact of Rewards}
\label{exp:imp_of_reward}
We examine the impact of reward on the equilibrium points in the CEF by analyzing the effect of $r_1$ on the payoff defined in \eq{homo-pure-payoff}.  
%
As shown in \cref{figr1}, the probability that EIP $1$ contributes $4$ workers increases with $r_1$. 
Similarly, EIP $2$ increases the probability of contributing $8$ workers to the CEF. This can be explained by the increased incentive provided by the CEF to the EIPs. 

Figure \ref{figr2} shows that the average utility received by each EIP increases as the reward issued by the CEF increases.
In addition, the increase to EIP $2$ is larger since on average EIP $2$ with $L_2=8$ invests more workers than EIP $1$ with $L_1=4$ to the CEF. 

\subsubsection{Impact of the Code Configuration}
\label{exp:imp_of_code}

This section examines the impact of the code configuration on the equilibrium points in the CEF. Figure \ref{figevolve} shows that all fractional evolutionary games with different values of $\alpha$ can converge to the same equilibrium point. Thus, we only plot the case when $\alpha=1$. 
Figure \ref{fign1} shows that the probability of choosing the last strategy, i.e., investing $4$ workers, for EIP $1$ increases as $n$ increases. A similar trend is observed for the EIP $2$.
The reason is that as the number of workers required by the TP increases, it is more likely that EIP $1$ must invest its maximum number of workers to meet the TP's requirement and to earn the reward. Thus, we see an increasing trend of population states as $n$ increases. 
Note that the maximum total contribution by both EIPs is $12$ workers. Consequently, when the value of $n$, which is part of the TP's request, is greater than $12$, the federation begins to fail to meet the TP's requirement and receives no reward. Consequently, we see that both EIPs begin to reduce their contributions.

On the other hand, \cref{fign2} shows that the population states of the last strategy in equilibrium increases as $k$ increases, because with more workers contributed to the CEF, the EIPs are more likely to return results in the first $k$ positions and thus receive higher additional rewards.

\section{Conclusion}
\label{sec conclusion}
In this paper, we proposed coded edge federation (CEF) to address the problem of insufficient edge resources when implementing CDC services in edge networks. In particular, we adopted an evolutionary game approach to model the dynamic behaviors of EIPs. To enhance the classical replicator dynamics, we proposed fractional replicator dynamics that incorporates a power-law fading memory via the left-sided Caputo derivatives. We theoretically showed and experimentally verified the validity of the fractional replicator dynamics in the sense that the equilibrium point of the game exists and is evolutionarily stable. In addition, the experiments demonstrate that the fractional replicator dynamics can model EIPs with additional features, e.g., the sensitivity or aggressiveness of the strategy adaptation and can provide a faster convergence rate than the classical one. The equilibrium adaptation for different hyper-parameter values is also investigated. The incorporation of pricing models for the EIP resources will be considered in future work. 
\section*{Acknowledgment}
The authors would like to thank...

\ifCLASSOPTIONcaptionsoff
  \newpage
\fi

\bibliographystyle{IEEEtran}
\bibliography{all.bib}

\begin{thebibliography}{10}
\providecommand{\url}[1]{#1}
\csname url@samestyle\endcsname
\providecommand{\newblock}{\relax}
\providecommand{\bibinfo}[2]{#2}
\providecommand{\BIBentrySTDinterwordspacing}{\spaceskip=0pt\relax}
\providecommand{\BIBentryALTinterwordstretchfactor}{4}
\providecommand{\BIBentryALTinterwordspacing}{\spaceskip=\fontdimen2\font plus
\BIBentryALTinterwordstretchfactor\fontdimen3\font minus
  \fontdimen4\font\relax}
\providecommand{\BIBforeignlanguage}[2]{{%
\expandafter\ifx\csname l@#1\endcsname\relax
\typeout{** WARNING: IEEEtran.bst: No hyphenation pattern has been}%
\typeout{** loaded for the language `#1'. Using the pattern for}%
\typeout{** the default language instead.}%
\else
\language=\csname l@#1\endcsname
\fi
#2}}
\providecommand{\BIBdecl}{\relax}
\BIBdecl

\bibitem{leeSpeedingDistributedMachine2018a}
K.~Lee, M.~Lam, R.~Pedarsani, D.~Papailiopoulos, and K.~Ramchandran, ``Speeding
  {{Up Distributed Machine Learning Using Codes}},'' \emph{IEEE Transactions on
  Information Theory}, vol.~64, no.~3, pp. 1514--1529, Mar. 2018.

\bibitem{reisizadehCodedComputationHeterogeneous2017}
A.~Reisizadeh, S.~Prakash, R.~Pedarsani, and S.~Avestimehr, ``Coded computation
  over heterogeneous clusters,'' in \emph{{{IEEE International Symposium}} on
  {{Information Theory}}}, Jun. 2017, pp. 2408--2412.

\bibitem{yuPolynomialCodesOptimal2017}
Q.~Yu, M.~A. {Maddah-Ali}, and A.~S. Avestimehr, ``Polynomial codes: An optimal
  design for high-dimensional coded matrix multiplication,'' in
  \emph{Proceedings of the 31st {{International Conference}} on {{Neural
  Information Processing Systems}}}, ser. {{NIPS}}'17.\hskip 1em plus 0.5em
  minus 0.4em\relax {Curran Associates Inc.}, Dec. 2017, pp. 4406--4416.

\bibitem{tandonGradientCodingAvoiding2017}
R.~Tandon, Q.~Lei, A.~G. Dimakis, and N.~Karampatziakis,
  ``\BIBforeignlanguage{en}{Gradient {{Coding}}: {{Avoiding Stragglers}} in
  {{Distributed Learning}}},'' in \emph{\BIBforeignlanguage{en}{International
  {{Conference}} on {{Machine Learning}}}}.\hskip 1em plus 0.5em minus
  0.4em\relax {PMLR}, Jul. 2017, pp. 3368--3376.

\bibitem{ravivGradientCodingCyclic2019}
N.~Raviv, I.~Tamo, R.~Tandon, and A.~G. Dimakis, ``Gradient {{Coding From
  Cyclic MDS Codes}} and {{Expander Graphs}},'' \emph{IEEE Transactions on
  Information Theory}, vol.~66, no.~12, pp. 7475--7489, Dec. 2020.

\bibitem{yuLagrangeCodedComputing2019}
Q.~Yu, S.~Li, N.~Raviv, S.~M.~M. Kalan, M.~Soltanolkotabi, and S.~A.
  Avestimehr, ``\BIBforeignlanguage{en}{Lagrange {{Coded Computing}}: {{Optimal
  Design}} for {{Resiliency}}, {{Security}}, and {{Privacy}}},'' in
  \emph{\BIBforeignlanguage{en}{The 22nd {{International Conference}} on
  {{Artificial Intelligence}} and {{Statistics}}}}.\hskip 1em plus 0.5em minus
  0.4em\relax {PMLR}, Apr. 2019, pp. 1215--1225.

\bibitem{liCodedTeraSort2017}
S.~Li, S.~Supittayapornpong, M.~A. {Maddah-Ali}, and S.~Avestimehr,
  ``\BIBforeignlanguage{English}{Coded {{TeraSort}}},'' in
  \emph{\BIBforeignlanguage{English}{{{IEEE International Parallel}} and
  {{Distributed Processing Symposium}}: {{Workshops}} ({{IPDPSW}})}}.\hskip 1em
  plus 0.5em minus 0.4em\relax {IEEE Computer Society}, May 2017, pp. 389--398.

\bibitem{liCodedComputingMitigating2020}
S.~Li and S.~Avestimehr, ``\BIBforeignlanguage{English}{Coded {{Computing}}:
  {{Mitigating Fundamental Bottlenecks}} in {{Large}}-{{Scale Distributed
  Computing}} and {{Machine Learning}}},''
  \emph{\BIBforeignlanguage{English}{Now Foundations and Trends}}, 2020.

\bibitem{ngComprehensiveSurveyCoded2021}
J.~S. Ng, W.~Y.~B. Lim, N.~C. Luong, Z.~Xiong, A.~Asheralieva, D.~Niyato,
  C.~Leung, and C.~Miao, ``A {{Comprehensive Survey}} on {{Coded Distributed
  Computing}}: {{Fundamentals}}, {{Challenges}}, and {{Networking
  Applications}},'' \emph{IEEE Communications Surveys \& Tutorials}, vol.~23,
  no.~3, pp. 1800--1837, 2021.

\bibitem{prakashCodedComputingLowLatency2021}
S.~Prakash, S.~Dhakal, M.~Akdeniz, Y.~Yona, S.~Talwar, S.~Avestimehr, and
  N.~Himayat, ``Coded {{Computing}} for {{Low}}-{{Latency Federated Learning}}
  over {{Wireless Edge Networks}},'' \emph{IEEE Journal on Selected Areas in
  Communications}, vol.~39, no.~1, pp. 233--250, Jan. 2021.

\bibitem{haCodedFederatedComputing2019}
S.~Ha, J.~Zhang, O.~Simeone, and J.~Kang, ``Coded {{Federated Computing}} in
  {{Wireless Networks}} with {{Straggling Devices}} and {{Imperfect CSI}},'' in
  \emph{{{IEEE International Symposium}} on {{Information Theory}}}, Jul. 2019,
  pp. 2649--2653.

\bibitem{kimCodedEdgeComputing2020}
K.~T. Kim, C.~{Joe-Wong}, and M.~Chiang, ``Coded {{Edge Computing}},'' in
  \emph{{{IEEE INFOCOM}}}, Jul. 2020, pp. 237--246.

\bibitem{hanOpportunisticCodedDistributed2021}
Y.~Han, D.~Niyato, C.~Leung, and D.~I. Kim, ``Opportunistic {{Coded Distributed
  Computing}}: {{An Evolutionary Game Approach}},'' in \emph{International
  {{Wireless Communications}} and {{Mobile Computing}} ({{IWCMC}})}, Jun. 2021,
  pp. 1430--1435.

\bibitem{wangCloudComputingPerspective2010}
L.~Wang, G.~Von~Laszewski, A.~Younge, X.~He, M.~Kunze, J.~Tao, and C.~Fu,
  ``Cloud computing: A perspective study,'' \emph{New generation computing},
  vol.~28, no.~2, pp. 137--146, 2010.

\bibitem{rashidDistributedCloudComputing2018}
Z.~N. Rashid, S.~R. Zebari, K.~H. Sharif, and K.~Jacksi, ``Distributed cloud
  computing and distributed parallel computing: {{A}} review,'' in
  \emph{International {{Conference}} on {{Advanced Science}} and
  {{Engineering}} ({{ICOASE}})}.\hskip 1em plus 0.5em minus 0.4em\relax {IEEE},
  2018, pp. 167--172.

\bibitem{agarwalVolleyAutomatedData2010}
S.~Agarwal, J.~Dunagan, N.~Jain, S.~Saroiu, A.~Wolman, and H.~Bhogan, ``Volley:
  Automated data placement for geo-distributed cloud services,'' in
  \emph{Proceedings of the 7th {{USENIX}} Conference on {{Networked}} Systems
  Design and Implementation}, ser. {{NSDI}}'10.\hskip 1em plus 0.5em minus
  0.4em\relax {USA}: {USENIX Association}, Apr. 2010, p.~2.

\bibitem{alicherryNetworkAwareResource2012}
M.~Alicherry and T.~V. Lakshman, ``Network aware resource allocation in
  distributed clouds,'' in \emph{Proceedings {{IEEE INFOCOM}}}.\hskip 1em plus
  0.5em minus 0.4em\relax {IEEE}, 2012, pp. 963--971.

\bibitem{endoResourceAllocationDistributed2011}
P.~T. Endo, A.~V. {de Almeida Palhares}, N.~N. Pereira, G.~E. Goncalves,
  D.~Sadok, J.~Kelner, B.~Melander, and J.-E. Mangs, ``Resource allocation for
  distributed cloud: Concepts and research challenges,'' \emph{IEEE Network},
  vol.~25, no.~4, pp. 42--46, 2011.

\bibitem{coadyDistributedCloudComputing2015}
Y.~Coady, O.~Hohlfeld, J.~Kempf, R.~McGeer, and S.~Schmid, ``Distributed cloud
  computing: {{Applications}}, status quo, and challenges,'' \emph{ACM SIGCOMM
  Computer Communication Review}, vol.~45, no.~2, pp. 38--43, 2015.

\bibitem{caoEdgeFederationIntegrated2020}
X.~Cao, G.~Tang, D.~Guo, Y.~Li, and W.~Zhang, ``Edge {{Federation}}:
  {{Towards}} an {{Integrated Service Provisioning Model}},'' \emph{IEEE/ACM
  Transactions on Networking}, vol.~28, no.~3, pp. 1116--1129, Jun. 2020.

\bibitem{liFundamentalTradeoffComputation2018}
S.~Li, M.~A. {Maddah-Ali}, Q.~Yu, and A.~S. Avestimehr, ``A {{Fundamental
  Tradeoff Between Computation}} and {{Communication}} in {{Distributed
  Computing}},'' \emph{IEEE Transactions on Information Theory}, vol.~64,
  no.~1, pp. 109--128, Jan. 2018.

\bibitem{ngGametheoreticApproachCollaborative2021}
J.~S. Ng, W.~Y.~B. Lim, Z.~Xiong, D.~Niyato, C.~Leung, D.~I. Kim, J.~Zhang, and
  Q.~Yang, ``A {{Game}}-theoretic {{Approach Towards Collaborative Coded
  Computation Offloading}},'' \emph{arXiv:2102.08667 [cs]}, Feb. 2021.

\bibitem{kimIncentiveBasedCodedDistributed2020}
N.~Kim, D.~Kim, J.~Lee, D.~Niyato, and J.~K. Choi,
  ``\BIBforeignlanguage{en}{Incentive-{{Based Coded Distributed Computing
  Management}} for {{Latency Reduction}} in {{IoT Services}} \textendash{{A
  Game Theoretic Approach}}},'' \emph{\BIBforeignlanguage{en}{IEEE Internet of
  Things Journal}}, pp. 8259--8278, 2020.

\bibitem{weibullEvolutionaryGameTheory1997}
J.~W. Weibull, \emph{Evolutionary Game Theory}.\hskip 1em plus 0.5em minus
  0.4em\relax {MIT press}, 1997.

\bibitem{tarasovaConceptDynamicMemory2018}
V.~V. Tarasova and V.~E. Tarasov, ``\BIBforeignlanguage{en}{Concept of dynamic
  memory in economics},'' \emph{\BIBforeignlanguage{en}{Communications in
  Nonlinear Science and Numerical Simulation}}, vol.~55, pp. 127--145, Feb.
  2018.

\bibitem{kilbasTheoryApplicationsFractional2006}
A.~A. Kilbas, H.~M. Srivastava, and J.~J. Trujillo, \emph{Theory and
  Applications of Fractional Differential Equations}.\hskip 1em plus 0.5em
  minus 0.4em\relax {Elsevier Science}, 2006, vol. 204.

\bibitem{parkHierarchicalCodingDistributed2018}
H.~Park, K.~Lee, J.-Y. Sohn, C.~Suh, and J.~Moon, ``Hierarchical {{Coding}} for
  {{Distributed Computing}},'' in \emph{{{IEEE International Symposium}} on
  {{Information Theory}}}, Jun. 2018, pp. 1630--1634.

\bibitem{bitarMinimizingLatencySecure2020}
R.~Bitar, P.~Parag, and S.~El~Rouayheb, ``Minimizing {{Latency}} for {{Secure
  Coded Computing Using Secret Sharing}} via {{Staircase Codes}},'' \emph{IEEE
  Transactions on Communications}, vol.~68, no.~8, pp. 4609--4619, Aug. 2020.

\bibitem{wangJointcloudCrosscloudCooperation2017}
H.~Wang, P.~Shi, and Y.~Zhang, ``{{JointCloud}}: {{A Cross}}-{{Cloud
  Cooperation Architecture}} for {{Integrated Internet Service
  Customization}},'' in \emph{{{IEEE}} 37th {{International Conference}} on
  {{Distributed Computing Systems}} ({{ICDCS}})}, Jun. 2017, pp. 1846--1855.

\bibitem{meylerMicrosoftHybridCloud2017}
K.~Meyler, S.~Buchanan, M.~Scholman, J.~G. Svendsen, and J.~Rangama,
  \emph{Microsoft {{Hybrid Cloud Unleashed}} with {{Azure Stack}} and
  {{Azure}}}.\hskip 1em plus 0.5em minus 0.4em\relax {Sams Publishing}, 2017.

\bibitem{dziyauddinComputationOffloadingContent2019}
R.~A. Dziyauddin, D.~Niyato, N.~C. Luong, M.~A.~M. Izhar, M.~Hadhari, and
  S.~Daud, ``Computation {{Offloading}} and {{Content Caching Delivery}} in
  {{Vehicular Edge Computing}}: {{A Survey}},'' \emph{arXiv:1912.07803 [cs]},
  Dec. 2019.

\bibitem{shanMultilevelOptimizationFramework2020}
N.~Shan, Y.~Li, and X.~Cui, ``\BIBforeignlanguage{en}{A {{Multilevel
  Optimization Framework}} for {{Computation Offloading}} in {{Mobile Edge
  Computing}}},'' \emph{\BIBforeignlanguage{en}{Mathematical Problems in
  Engineering}}, vol. 2020, Jun. 2020.

\bibitem{danDynamicContentAllocation2014}
G.~D{\'a}n and N.~Carlsson, ``Dynamic content allocation for cloud-assisted
  service of periodic workloads,'' in \emph{{{IEEE INFOCOM}} 2014-{{IEEE
  Conference}} on {{Computer Communications}}}.\hskip 1em plus 0.5em minus
  0.4em\relax {IEEE}, 2014, pp. 853--861.

\bibitem{koellerApplicationsFractionalCalculus1984}
R.~C. Koeller, ``Applications of {{Fractional Calculus}} to the {{Theory}} of
  {{Viscoelasticity}},'' \emph{Journal of Applied Mechanics}, vol.~51, no.~2,
  pp. 299--307, Jun. 1984.

\bibitem{colemanGeneralTheoryFading1968}
B.~D. Coleman and V.~J. Mizel, ``On the general theory of fading memory,''
  \emph{Archive for Rational Mechanics and Analysis}, vol.~29, no.~1, pp.
  18--31, 1968.

\bibitem{teyssiereLongMemoryEconomics2006}
G.~Teyssi{\`e}re and A.~P. Kirman, \emph{Long Memory in Economics}.\hskip 1em
  plus 0.5em minus 0.4em\relax {Springer Science \& Business Media}, 2006.

\bibitem{tarasovaElasticityEconomicProcesses2016}
V.~V. Tarasova and V.~E. Tarasov, ``Elasticity for economic processes with
  memory: {{Fractional}} differential calculus approach,'' \emph{Fractional
  Differential Calculus}, vol.~6, no.~2, pp. 219--232, 2016.

\bibitem{podlubnyFractionalDifferentialEquations1998}
I.~Podlubny, \emph{Fractional Differential Equations: An Introduction to
  Fractional Derivatives, Fractional Differential Equations, to Methods of
  Their Solution and Some of Their Applications}.\hskip 1em plus 0.5em minus
  0.4em\relax {Elsevier}, 1998.

\bibitem{tarasovaEconomicGrowthModel2017}
V.~Tarasova and V.~Tarasov, ``Economic {{Growth Model}} with {{Constant Pace}}
  and {{Dynamic Memory}},'' \emph{Problems of modern science and education},
  vol.~84, Jan. 2017.

\bibitem{tongHierarchicalEdgeCloud2016}
L.~Tong, Y.~Li, and W.~Gao, ``A hierarchical edge cloud architecture for mobile
  computing,'' in \emph{{{IEEE INFOCOM}}}, Apr. 2016, pp. 1--9.

\bibitem{haoOnlineAllocationVirtual2017}
F.~Hao, M.~Kodialam, T.~V. Lakshman, and S.~Mukherjee, ``Online {{Allocation}}
  of {{Virtual Machines}} in a {{Distributed Cloud}},'' \emph{IEEE/ACM
  Transactions on Networking}, vol.~25, no.~1, pp. 238--249, Feb. 2017.

\bibitem{chandyAnalysisResourceCosts2009}
J.~A. Chandy, ``\BIBforeignlanguage{en}{An analysis of resource costs in a
  public computing grid},'' in \emph{\BIBforeignlanguage{en}{{{IEEE
  International Symposium}} on {{Parallel}} \& {{Distributed
  Processing}}}}.\hskip 1em plus 0.5em minus 0.4em\relax {Rome, Italy}: {IEEE},
  May 2009, pp. 1--8.

\bibitem{sandholmPopulationGamesEvolutionary2010}
W.~H. Sandholm, \emph{Population Games and Evolutionary Dynamics}.\hskip 1em
  plus 0.5em minus 0.4em\relax {MIT press}, 2010.

\bibitem{niyatoDynamicsNetworkSelection2008}
D.~Niyato and E.~Hossain, ``Dynamics of network selection in heterogeneous
  wireless networks: {{An}} evolutionary game approach,'' \emph{IEEE
  Transactions on Vehicular Technology}, vol.~58, no.~4, 2008.

\bibitem{jiangGraphicalEvolutionaryGame2014}
C.~Jiang, Y.~Chen, and K.~R. Liu, ``Graphical evolutionary game for information
  diffusion over social networks,'' \emph{IEEE Journal of Selected Topics in
  Signal Processing}, vol.~8, no.~4, pp. 524--536, 2014.

\bibitem{dongJointOptimizationTask2019c}
C.~Dong and W.~Wen, ``Joint optimization for task offloading in edge computing:
  {{An}} evolutionary game approach,'' \emph{Sensors}, vol.~19, no.~3, p. 740,
  2019.

\bibitem{cuiNovelMethodMobile2020}
Y.~Cui, D.~Zhang, T.~Zhang, L.~Chen, M.~Piao, and H.~Zhu, ``Novel method of
  mobile edge computation offloading based on evolutionary game strategy for
  {{IoT}} devices,'' \emph{AEU-International Journal of Electronics and
  Communications}, vol. 118, 2020.

\bibitem{daniellTheoryFunctionalsIntegral1932}
P.~J. Daniell, ``\BIBforeignlanguage{en}{The {{Theory}} of {{Functionals}}, and
  of {{Integral}} and {{Integro}}-differential {{Equations}}. {{By V}}.
  {{Volterra}}. {{Translated}} by {{M}}. {{Long}}. {{Pp}}. x + 226. 25s. 1930.
  ({{Blackie}})},'' \emph{\BIBforeignlanguage{en}{The Mathematical Gazette}},
  vol.~16, no. 217, pp. 59--60, Feb. 1932.

\bibitem{edelmanEvolutionSystemsPowerLaw2020}
M.~Edelman, ``Evolution of {{Systems}} with {{Power}}-{{Law Memory}}: {{Do We
  Have}} to {{Die}}?({{Dedicated}} to the {{Memory}} of {{Valentin
  Afraimovich}}),'' in \emph{Demography of {{Population Health}}, {{Aging}} and
  {{Health Expenditures}}}.\hskip 1em plus 0.5em minus 0.4em\relax {Springer},
  2020, pp. 65--85.

\bibitem{andersonLearningMemoryIntegrated2000}
J.~R. Anderson, \emph{Learning and Memory: {{An}} Integrated Approach}.\hskip
  1em plus 0.5em minus 0.4em\relax {John Wiley \& Sons Inc}, 2000.

\bibitem{wixtedAnalyzingEmpiricalCourse1990}
J.~T. Wixted, ``Analyzing the empirical course of forgetting.'' \emph{Journal
  of Experimental Psychology: Learning, Memory, and Cognition}, vol.~16, no.~5,
  pp. 927--935, 1990.

\bibitem{diethelmAnalysisFractionalDifferential2002a}
K.~Diethelm and N.~J. Ford, ``Analysis of fractional differential equations,''
  \emph{Journal of Mathematical Analysis and Applications}, vol. 265, no.~2,
  pp. 229--248, 2002.

\bibitem{daftardar-gejjiAnalysisSystemNonautonomous2007}
V.~{Daftardar-Gejji} and H.~Jafari, ``Analysis of a system of nonautonomous
  fractional differential equations involving {{Caputo}} derivatives,''
  \emph{Journal of Mathematical Analysis and Applications}, vol. 328, no.~2,
  pp. 1026--1033, 2007.

\bibitem{baleanuGlobalExistenceSolutions2010}
D.~B{\u a}leanu and O.~G. Mustafa, ``On the global existence of solutions to a
  class of fractional differential equations,'' \emph{Computers \& Mathematics
  with Applications}, vol.~59, no.~5, pp. 1835--1841, 2010.

\bibitem{el-raheemModificationApplicationContraction2003}
Z.~F. {El-Raheem}, ``Modification of the application of a contraction mapping
  method on a class of fractional differential equation,'' \emph{Applied
  mathematics and computation}, vol. 137, no. 2-3, pp. 371--374, 2003.

\bibitem{abdel-salamStabilityFractionalorderNonautonomous2007}
S.~A. {AbdEl-Salam} and A.~{El-Sayed}, ``On the stability of some
  fractional-order non-autonomous systems,'' \emph{Electronic Journal of
  Qualitative Theory of Differential Equations}, vol. 2007, no.~6, pp. 1--14,
  2007.

\bibitem{sahooMeanValueTheorems1998}
P.~Sahoo and T.~Riedel, \emph{\BIBforeignlanguage{English}{Mean Value Theorems
  and Functional Equations}}.\hskip 1em plus 0.5em minus 0.4em\relax {World
  Scientific}, 1998.

\end{thebibliography}

\appendices
%

\section{Equivalent Norm}\label{app:norm}
\begin{proof}
Let us observe that for the norm $\norm{\cdot}^*$, the following inequalities are valid for any function $\bm{x}\in C(J,\bb{R}^L)$:
\begin{equation}
e^{-NT}\norm{\bm{x}}\leq \|\bm{x}\|^{*} \leq \sum_{\cond} \underset{t}{\sup} \left|x_{i,j} (t)\right|=\norm{\bm{x}},
\end{equation}
where we denote as $\norm{\cdot}$ the supremum norm on $C(J,\bb{R}^L)$. Here, $T=\max\{t;t\in J\}$. Thus, norms $\norm{\cdot}$ and $\norm{\cdot}^*$ are equivalent and so are the metrics. 
\end{proof}

\section{Lipschizian Conditions}
\label{appendix lipschiz}
\begin{proof}
To prove \cref{lip corollary}, we can show that \cref{simple lim} holds. If \cref{simple lim} holds, then \cref{lip corollary} can be proved straightforwardly by the triangle inequality. Note that in this proof, for simplicity, we re-index the component by letting $[i,j]_{\cond}\equiv [k]_{1\leq k\leq L}$.
\begin{multline} 
|\phi_{i,j}(x_{1,0}(t), \ldots,x_{h,k}(t), \ldots, x_{m,l_m+1}(t)) 
-\\
\phi_{i,j}(x_{1,0}(t), \ldots,y_{h,k}(t), \ldots, x_{m,l_m+1}(t))|  
\\\leqq A_{i,j}|x_{h,k}(t)-y_{h,k}(t)|
\quad (A_{i,j}>0 ; \\
i,h \in \mathcal{I},0\leq j\leq L_i,0\leq k\leq L_h).
\label{simple lim}
\end{multline}

To verify \cref{simple lim}, we can prove the partial derivative $\frac{\partial\phi_{i,j}}{\partial x_{h,k}}, (i,h \in \mathcal{I},0\leq j\leq L_i,0\leq k\leq L_h)$ exists and bounded on $\Theta$, i.e., $\exists M \ge0, s.t., |\frac{\partial\phi_{i,j}}{\partial x_{h,k}}|\leq M$. The reason is that with bounded partial derivatives, \cref{simple lim} is true following the Mean Value Theorem \cite{sahooMeanValueTheorems1998}.

The partial derivatives can be expressed as follows:
		\begin{align*}
			\left| \frac{\partial\phi_{i,j}}{\partial x_{h,k}}\right| 
			&= \left|\frac{\partial x_{i,j}}{\partial x_{h,k}}+x_{i,j}\left(\frac{\partial u_{i, j}(\mb{x})}{\partial x_{h,k}}-\frac{\partial {u}_{i}(\mb{x})}{\partial x_{h,k}}\right)\right| \\
			&\leq \left| \frac{\partial x_{i,j}}{\partial x_{h,k}}\right|  + \left| x_{i,j}\right| \left| \frac{\partial u_{i, j}(\mb{x})}{\partial x_{h,k}}\right| +\left| x_{i,j}\right| ^2\left| \frac{\partial {u}_{i,j}(\mb{x})}{\partial x_{h,k}}\right| .
		\end{align*}
It is easy to check that $\left| \frac{\partial x_{i,j}}{\partial x_{h,k}}\right| $ and $\left| x_{i,j}\right| $ are bounded, and therefore, it is only left to show that $\frac{\partial {u}_{i,j}(\mb{x})}{\partial x_{h,k}}$ is bounded. We only prove for the case of $2$ EIPs, which can be extended to more numbers of EIPs straightforwardly. 

1) If $h=-i$, $\left| \frac{\partial\phi_{i,j}}{\partial x_{h,k}}\right| $ is equivalent to verify that $\pi_i(j,q)$ is bounded, which is equivalent to show that $C_{i,j}^{(2)}$ is bounded as it is the only term involved the decision variables. Note that $\left| C_{i,j}^{(2)}\right| =\left| \frac{x_{i,j}j}{\sum_{q\in Q_i}{x_{i,q}q}} \frac{f(w_i)}{E_i}\right|\leq 
\left| \frac{x_{i,j}j}{E_i} f(w_i)\right|
\leq \left| \frac{j}{E_i} f(w_i) \right|
$. $f(w_i)$ is bounded as it reaches its maximum values when all the edge instances are configured to the maximum number of workers to the edge federation. 

2) if $h=i$, $\left| \frac{\partial\phi_{i,j}}{\partial x_{h,k}}\right| $ is equivalent to show that $\left| \frac{\partial C_{i,j}^{(2)}}{\partial x_{i,k}}\right|$ is bounded. Let $\zeta$ denote $\frac{f(w_i)}{\sum_{q \in Q_{i}} x_{i, q} q}$. Then, $\left| \frac{\partial C_{i,j}^{(2)}}{\partial x_{i,k}}\right|
\leq \left| \frac{\partial x_{i,j}}{\partial x_{i,h}} \right|\left| \frac{j}{E_i} \right|\left| \zeta \right|   +\left| \frac{1}{E_i} \right| \left| \frac{\partial \zeta}{\partial x_{i,k}}\right|
$. Here $\zeta$ is bounded, as $\left| \zeta \right| \leq \left| f(w_i) \right| $, and $f(w_i)$ is proved bounded in (1). Let $\tilde{q}=\max\{q;q\in Q_i\}$. Then $\left| \frac{\partial *}{\partial x_{i,k}} \right| 
\leq \left| \frac{\partial f(w_i)}{\partial x_{i,k}} \tilde{q} \right| +\left|k\right|\left| f(w_i)\right| $.
Here,
$\left| \frac{\partial f(w_i)}{\partial x_{i,k}}   \right| 
=\left| \frac{1}{(1-w_i)^2}\right| \left| \frac{E_ik}{W_i} \right| 
$, and 
$\frac{1}{(1-w_i)^2}$ is bounded and its maximum value reaches when $w_i\in (0,1)$ reaches its maximized values, i.e., when all edge instances configured to the maximum number of workers. 

Thus, $\left| \frac{\partial\phi_{i,j}}{\partial x_{h,k}}\right| $ is proved bounded, so \cref{simple lim} is bounded based on the Mean Value Theorem. 
\end{proof}

\section{Uniqueness and Existence}
\label{appendix unique}
\begin{proof}
The \cref{voltera} can be expressed as follows:
\begin{equation}
\bm{x}(t)=\bm{x}_o + I_{0+}^\alpha \phi(\bm{x}), \quad 0<\alpha<1 \label{integral problem}
\end{equation}
where $\bm{x}_o=[x_{i,j}(0)]_{\cond}$ denotes the vectored initial values. $\bm{x}(t)=[x_{i,j}(t)]_{\cond}$ and $\phi=[\phi_{i,j}(\bm{x}(t))]_{\cond},t\in J$ are defined previously. 


Let $K=\underset{k}{\max}\{A_k\}$, where $A_k$ is given in the definition of the Lipschizian condition, then we can have the below inequalities adapted from \cref{lip corollary},
\begin{align}
&\left|\phi_{i,j}\left(\bm{x}(t)\right)-\phi_{i,j}\left(\bm{y}(t)\right)\right|  \leq K \norm{\bm{x}(t)-\bm{y}(t)}_L,  \label{cond1}\\
&\left|\frac{\partial}{\partial x_{i,j}} \phi_{i,j}\right|\leq K \quad (\forall t\in J, \cond) , \label{cond2}
\end{align} 
where $|\cdot|$ is the absolute value and $\norm{\bm{x}}_L=\sum_{\cond}|x_{i,j}|$ on $\bb{R}^L$.

Let the operator $T: C(J,\bb{R}^L)\rightarrow C(J,\bb{R}^L)$ be defined by 
\begin{equation}
T\bx(t)=\bx_o+I^\alpha \phi(\bx(t))
\label{vector votera}
\end{equation}
with the norm $\norm{\bx}^*$. The fixed point under the mapping $T$, i.e., $T\bx=\bx$, is the stationary point to the original systems of fractional differential equations defined in \cref{ivp-eq,ivp-value}. Thus, it is the equilibrium point in the fractional evolutionary game. 

To show the \textit{existence} and \textit{uniqueness} of such a fixed point, the idea is to apply the Banach Fixed Point Theorem (See Chapter 1 in \cite{kilbasTheoryApplicationsFractional2006} for more details) over the space of $C(J,\bb{R}^L)$ with norm $\norm{\bx}^*$. Let $\bx(t)$ and $\bm{y}(t)$ be any two points in the vector space $C(J,\bb{R}^L)$. We want to show that the operator $T$ defined in \cref{vector votera} is a contractive mapping, i.e. $\norm{T\bm{x}-T\bm{y}}^*<\norm{\bm{x}-\bm{y}}^*$. 

First of all, we prove two inequalities as follows:

\begin{align}
A 
&=\sum_{\cond}e^{-Ns} |\phi_{i,j}(\bm{x}(s))-\phi_{i,j}(\bm{y}(s))| \nonumber \\
&\leq  \sum_{\cond} e^{-Ns}  K\norm{\bm{x}(s)-\bm{y}(s)}_L \quad (\cref{cond1})\nonumber\\ 
&= e^{-Ns} LK  \norm{\bm{x}(s)-\bm{y}(s)}_L \nonumber\\
&= LK \sum_{\cond} |x_i(s)-y_i(s)| e^{-Ns}\nonumber\\
&\leq LK \sum_{\cond} \underset{s}{\sup}|x_{i,j}(s)-y_{i,j}(s)| e^{-Ns}\nonumber\\
&=LK\norm{\bm{x}-\bm{y}}^{*}.
\end{align}

\begin{align}
B&=\int_{0}^{t}(t-s)^{\alpha-1} e^{-N(t-s)}   ds \nonumber \\
 &\stackrel{z=(t-s)}{=}\int_{0}^{t}z^{\alpha-1}e^{-Nz}dz \nonumber\\ &\stackrel{\tau=Nz}{=}\int_{0}^{Nt} \frac{1}{N^\alpha}\tau^{\alpha-1}e^{-\tau} d\tau \nonumber\\
 &<\frac{1}{N^\alpha} \underbrace{\int_{0}^{+\infty} \tau^{\alpha-1}e^{-\tau} d\tau}_{\Gamma{(\alpha)}}=\frac{\Gamma(\alpha) }{N^\alpha}.
\end{align}

Then, we have 
\begin{align}
&\norm{e^{-Nt}(T\bm{x}(t)-T\bm{y}(t))}_L \nonumber \\
&=\norm{e^{-N t}\{I^\alpha[\phi(\bm{x}(s))-\phi(\bm{y}(s))](t) \}}_L \nonumber\\
&=\norm{\frac{1}{\Gamma(\alpha)}\int_{0}^{t}(t-s)^{\alpha-1}e^{-N t} (\phi(\bm{x}(s))-\phi(\bm{y}(s)))ds}_L \nonumber\\
&\leq \frac{1}{\Gamma(\alpha)} \int_{0}^{t}(t-s)^{\alpha-1} e^{-N t}\norm{\phi(\bm{x}(s))-\phi(\bm{y}(s))}_L ds \nonumber\\
&\leq \frac{1}{\Gamma(\alpha)} \int_{0}^{t}(t-s)^{\alpha-1} e^{-N(t-s)}\underbrace{\norm{e^{-Ns}\phi(\bm{x}(s))-\phi(\bm{y}(s))}_L}_A  ds\nonumber \\
&\leq \frac{1}{\Gamma(\alpha)} \int_{0}^{t}(t-s)^{\alpha-1} e^{-N(t-s)}  LK\norm{\bm{x}-\bm{y}}^*  ds  \nonumber\\
&= LK\norm{\bm{x}-\bm{y}}^* \frac{1}{\Gamma(\alpha)} \underbrace{\int_{0}^{t}(t-s)^{\alpha-1} e^{-N(t-s)}   ds}_B\nonumber \\
&< \frac{LK}{N^\alpha}\norm{\bm{x}-\bm{y}}^* \quad \text{(Choosing $N$ s.t. $LK<N^\alpha$)}  
\label{inequality1} \\
&< \norm{\bm{x}-\bm{y}}^*.
\label{final inequality}
\end{align}

Then, we have
\begin{multline}
\norm{T\bx -  T\bm{y}}^*= \norm{ \underset{t}{\sup}\, e^{-Nt}(T\bm{x}(t)-T\bm{y}(t))}_L < \norm{\bm{x}-\bm{y}}^*.
\label{ieq}
\end{multline}
Here, the strict inequality in \cref{ieq} holds since the supremum is the same as the maximum value, since the mapping $e^{-Nt}(T\bx(t)-T\bm{y}(t))$ is continuous and defined on the closed interval $J$. 

Thus, operator $T$ is a contractive mapping. With the Banach Fixed Point Theorem, the operator $T$ has a unique fixed point. In other word, there exists a unique equilibrium point in the fractional evolutionary game.
\end{proof}

\section{Stability}
\label{appendix stability}
\begin{proof}
Let $\bx(t)$ be the solution to the initial value problem defined in \cref{ivp-eq,ivp-value} and $\bm{y}(t)$ be the solutions to the initial value problem given in \cref{ivp-eq,new initial value}. Following the notation in \cref{ivp-value,new initial value}, we use $\bm{x}_o \text{ and }\bm{y}_o$ to denote the corresponding initial points. 
Note that the solution $\bm{x}(t)\text{ and }\bm{y}(t)$ are the fixed point under its corresponding contractive mapping $T_x \text{ and } T_y$, where $(T_x\bm{x})(t)={\bx}_o+I^\alpha \phi(\bx(t))$ and $(T_y\bm{y})(t)=\bm{y}_o+I^\alpha \phi(\bm{y}(t))$. 
Then, we have
\begin{multline}
\norm{e^{-Nt}(\bm{x}-\bm{y})}_L=\norm{e^{-Nt}\{(T_x\bm{x})(t)-(T_y\bm{y})(t)\}}_L\\
=\norm{e^{-Nt}\{\bm{x}_o-\bm{y}_o +I^\alpha[\phi(\bm{x}(s))-\phi(\bm{y}(s))](t)\}}_L\\
\leq \underbrace{\norm{e^{-Nt}(\bm{x}_o-\bm{y}_o)}_L}_A  +
\underbrace{\norm{e^{-Nt}I^\alpha[\phi(\bm{x}(s))-\phi(\bm{y}(s))](t)}_L}_B
\label{proof2 stop1}
\end{multline}
where 
$
A=e^{-Nt}\norm{\bm{x}_o-\bm{y}_o}_L<\norm{\bm{x}_o-\bm{y}_o}_L,
$
as $N>0,t>0 \Rightarrow e^{-Nt}\in (0,1)$, and
$
B=e^{-Nt\{(T\bm{x})(t)-(T\bm{y})(t)\}}_L < \frac{l K}{N^{\alpha}}\|\bm{x}-\bm{y}\|^{*}, \forall t \in J$ as \cref{inequality1} holds. Thus, we have
\begin{align*}
&\norm{e^{-Nt}(\bm{x}-\bm{y})}_L < \norm{\bm{x}_o-Y_o}_L+\frac{l K}{N^{\alpha}}\|\bm{x}-\bm{y}\|^{*}  \quad \forall t\in J\\
&\Rightarrow  \norm{ \underset{t}{\sup} e^{-Nt}(\bm{x}-\bm{y})}_L < \norm{\bm{x}_o-\bm{y}_o}_L+\frac{l K}{N^{\alpha}}\|\bm{x}-\bm{y}\|^{*} \\
&\Rightarrow
\norm{\bm{x}-\bm{y}}^* < \norm{\bm{x}_o-\bm{y}_o}_L+\frac{l K}{N^{\alpha}}\|\bm{x}-\bm{y}\|^{*}\\
&\Rightarrow
\norm{\bm{x}-\bm{y}}^*  < \frac{1}{1-\frac{LK}{N^\alpha}} \norm{\bm{x}_o-\bm{y}_o}_L\\
&\Rightarrow
\norm{\bm{x}-\bm{y}}^*  < \norm{\bm{x}_o-\bm{y}_o}_L,
\end{align*} 
if $N$ is chosen the same as that in the \cref{main theorem}, which leads to  $\frac{1}{1-\frac{LK}{N^\alpha}}\in (0,1)$. Let $\delta=\epsilon$, then we can deduce that the equilibrium of the fractional evolutionary game defined in \cref{ivp-eq,ivp-value} are uniformly stable based on \cref{def stabilty}.
\end{proof}

\end{document}